\documentclass[12pt,preprint]{aastex}

\slugcomment{Accepted for publication in The Astronomical Journal, Apr 3, 2007; to appear in the July 2007 issue}

\def\iac{\'{\i}}
\begin{document}

\title{$U~B~V~R~I$ Photometry of Stellar Structures throughout the Disk of the Barred Galaxy NGC 3367}

\author{J. Antonio Garc{\iac}a-Barreto, H\'ector Hern\'andez-Toledo, Edmundo Moreno-D{\iac}az}
\affil{Instituto de Astronom{\iac}a,
Universidad Nacional Aut\'onoma de M\'exico, Apartado Postal 70-264,
M\'exico D.F. 04510 M\'exico}
\bigskip

\author{Tula Bernal-Mar{\iac}n and A. Luc{\iac}a Villarreal-Castillo}
\affil{Facultad de Ciencias,
Universidad Nacional Aut\'onoma de M\'exico,
M\'exico D.F. 04510 M\'exico}

\begin{abstract}

We report new detailed surface $U, B, V, R,$ and $I$ photometry of 81 stellar structures in the disk of the barred galaxy NGC 3367. The images show many different structures indicating that star formation is going on in the most part of the disk. NGC 3367 is known to have a very high concentration of molecular gas distribution in the central regions of the galaxy and bipolar synchrotron emission from the nucleus with two lobes (at 6 kpc) forming a triple structure similar to a radio galaxy. We have determined the $U, B, V, R,$ and $I$ magnitudes and $U - B~, B - V~, U - V~,$ and $V - I$ colors for the central region (nucleus), a region which includes supernovae 2003 AA, and 79 star associations throughout NGC 3367. Estimation of ages of star associations is very difficult due to several factors, among them: filling factor, metallicity, spatial distribution of each structure and the fact that we estimated the magnitudes with a circular aperture of 16 pixels in diameter, equivalent to $6''.8\sim1.4$ kpc. However, if the colors derived for NGC 3367 were similar to the colors expected of star clusters with theoretical evolutionary star tracks developed for the LMC and had a similar metallicity, NGC 3367 show 51\% of the observed structures with age type SWB I (few tens of Myrs), with seven sources outside the bright surface brightness visible disk of NGC 3367.

\end{abstract}

\section{Introduction}

        It is a continuous topic of study how a galaxy evolves with time \citep{lar77,van89,pfe00}. Integrated photometry of star clusters and star associations in a galaxy can provide fundamental insight on the star-formation history \citep{bic96}. $U, B, V$ colors of galaxies depend mainly on their history of star formation \citep{sea73,lar78}. The fact that morphologically peculiar galaxies show bluer $U - B$ colors in the two color diagram suggests that galaxies with peculiar appearance have experienced anomalous star formation histories
characterized by recent burst. Models with recent bursts of star formation (SFR) predict colors that in general lie off the normal relation for models with monotonically decreasing SFR \citep{lar78}, for example, a very recent strong burst results in an excessively blue $U - B$ for a given $B - V$ \citep{lar78}. Integrated photometry of star clusters and star associations in a galaxy would provide fundamental insight on the star-formation history \citep{bic96}.

    NGC~3367, the disk galaxy with a stellar bar of our study, is noteworthy because of
observational characteristics, some of them are: {\bf 1)} it shows a bipolar synchrotron outflow from the nucleus with a compact unresolved source with deconvolved diameter less than 65 pc, and two large lobes straddling the nucleus with a total, projected extent on the sky, from the north-east to the south-west lobe  of about 12 kpc (an image like a radio galaxy, but from, a nearby late disk galaxy); only the south-west lobe shows polarized emission, suggesting that this lobe is out of the plane of the disk and is closer to the observer than the other lobe \citep{gar98,gar02}; {\bf 2)} the axis of the ejected outflow is highly inclined with respect to the axis of rotation of the disk of the galaxy \citep{gar98,gar02}; {\bf 3)} there is a lot of molecular gas mostly concentrated in the central 9$''$ with an unresolved source with M(H$_2)\sim 5.9\times10^8$ M$_{\odot}$; this molecular mass suggests an optical extinction value A$_v(l)\sim75$ magnitudes toward the nucleus \citep{gar05} (as it has been detected through its CO(J=1-0) emission with the OVRO mm Array$\footnote{The Owens Valley Radio Observatory millimeter array is operated by Caltech
and it is supported by the National Science Foundation grant AST 99-81546}$); its total molecular (H$_2$) mass within 11.4 kpc diameter is M(H$_2)\sim 2.6\times10^9$ M$_{\odot}$ \citep{gar05}; {\bf 4)} it shows a bright southwest optical structure (resembling a ``bow shock'') with a collection of H$\alpha$ knots at a galactocentric radius of about 10 kpc from the nucleus \citep{san61,gar96a,gar96b}, {\bf 5)} it has {\it crooked} arms in its northwest side, and {\bf 6)} it seems to be isolated with no similar-diameter galaxy in its proximity; NGC 3367 lies, behind the Leo group of galaxies, at a distance of 43.6 Mpc, with a closest companion lying at a projected distance of $\approx$ 563 kpc or 18 optical diameters away \citep{gar03}.

    NGC~3367 is an SBc(s) barred spiral galaxy with a stellar bar structure of diameter $\approx 32''\sim 6.7$ kpc oriented at a position angle (P.A.) $\approx65--70^{\circ}$. The disk is inclined with respect to the plane of the sky at $\sim~30^{\circ}$ \citep{gar01}. NGC3367 has an X~ray luminosity much larger than any normal spiral galaxy \citep{gio90,sto91,fab92}. Its radio continuum  emission at a $4''.5$ angular resolution shows weak radio continuum emission extended throughout the disk \citep{gar98}. From Fabry Perot observations of the H$\alpha$ emission it is deduced that the kinematic major axis lies projected on the disk at a P.A. of 51$^{\circ}$ \citep{gar01}. Its optical spectrum shows moderately broad H$\alpha$+[NII] lines with FWHM$\sim650$ km s$^{-1}$, but its emission of H$\beta$ is stronger than its emission of [OIII]$\lambda 5007$\AA~ and there is weak emission of He II $\lambda 4686$\AA~   suggesting the existence of WR stars \citep{ver86,ver97,ho97}. NGC 3367 has been clasiffied as an HII nucleus (from the optical line ratios; see \citep{ho97a} and normal content of atomic hydrogen with M$_{HI}\sim~7\times10^9$ M$_{\odot}$ \citep{huc85}. Table 1 lists relevant information for NGC 3367 coming from the literature.

    Previous $U, B, V, I$ phometry of NGC 3367 has been obtained for the whole galaxy with magnitudes reported in RC3 \citep{vau93} and compiled in the NASA Extragalactic Data Base (NED)$\footnote{NED is operated by the Jet Propulsion Laboratory, California Institute of Technology under contract with the National Aeronautics and Space Administration.}$ as listed in Table 1; previously reported colors are as follows: $U - B = -0.16, ~B - V = 0.55, B - I$ = 1.45. Larson \& Tinsley (1978) have discussed different factors that would affect the $U, B, V$ colors, among them are {\it reddening} (galactic and intrinsic), {\it gaseous emission lines and nonthermal emission}, {\it chemical composition} and {\it initial mass function} (IMF). The color $B - V$ of NGC 3367 agrees with the color expected from a late type disk galaxy \citep{all76,lar78}; 
however the $U - B$ color of NGC 3367 certainly lie above the {\it mean} value of normal disk galaxies and the position in the $U - B$ vs $B - V$ \citep{lar78} is more in agreement with 
the statistical colors of galaxies from the Arp Atlas of peculiar galaxies $\footnote{As a comparison the $U - B$ and $B - V$ of an early type (elliptical) galaxy are $U - B\geq+0.55$, 
and $B - V\geq+0.80$ \citep{lar78}.}$. These colors of NGC 3367 suggest that this late type barred galaxy is currently having many regions with recent star formation.
\indent

    Although galactic reddening is a major source of uncertainity in the $U, B, V$ colors, for this galaxy it is not a major concern given its lalitude, $b=+58^{\circ}$. We report our magnitude estimates for the different structures corrected for Galactic extinction. Gaseous emission lines, different chemical composition and different initial mass function may contribute to the uncertainities of the expected colors \citep{lar78}.

\indent

    This paper reports new $U, B, V, R, I$ photometry of 81 individual structures throughout NGC 3367 using the 84 cm optical telescope in San Pedro M\'artir, Baja
California, M\'exico \citep{rich05}. For the data reduction and analysis we used the IRAF$\footnote{The IRAF package is written and supported by the IRAF programming group at the National Optical Astronomy Observatories (NOAO) in Tucson, Arizona. NOAO is operated by the Association of Universities for Research in Astronomy (AURA), Inc. under cooperative agreement with the National Science Foundation (NSF).}$ task {\it phot} with an aperture radius of 8 pixels ($3''.41\sim715$ pc). The existence of a close connection between violent galactic dynamical phenomena is consistent with theoretical expectations that high velocity collisions and shock fronts should be effective in compressing gas to high densities and triggering rapid star formation. Section 2 reports the observations, \S 3 reports the results and \S 4 gives the discussion and finally \S 5 describes the conclusions.

\section{Observations and Data Reduction}

    The observations were carried out using the 84 cm optical telescope at the Observatorio Astron\'omico Nacional at San Pedro M\'artir, Baja California, M\'exico on 2005, March 8 and 9. We used a CCD camera Site3 with a 1024$\times1024$ pixels; each pixel size is equivalent to $0''.426$ and we did not use any binning. Attached to the camera was a filter box. A journal of the photometric observations is given in Table 2. Column (1) give a characteristic; columns (2)-(6) give the filters; rows indicate the data for each column, seeing, central wavelength and total width of the filter used, and the number of frames times the integration time (in seconds) of each of the images (M67 and NGC 3367) in each filter. Images were debiased, trimmed, and flat-fielded using standard IRAF procedures. First, the bias level of the CCD was subtracted from all exposures. A run of 10 bias images was obtained per night, and these were combined into a single bias frame which was then applied to the object frames. The images were flat-fielded using sky flats taken in each filter at the beginning and at the end of each night. In the case of estimating the magnitudes for the structures throughout the disk in NGC 3367, we have made careful analysis utilizing different IRAF tasks, in particular, {\it qphot} and {\it phot} using different aperture diameters for bright stars in the field of NGC 3367 in order to estimate the uncertainities in the final magnitudes at different filters before estimating the final magnitudes reported for the different structures in NGC 3367. The magnitudes reported here for the structures within NGC 3367 were estimated using an aperture radius of 8 pixels ($\sim3''.4\sim715$ pc).
\indent

    Photometric calibration was achieved by nightly observations of standard stars of known magnitudes from the ``Dipper Asterism'' M67 star cluster \citep{gui91,che91}. A total of 9 standard stars with a color interval $ -0.1 \leq (B - V) \leq 1.4$ and a similar range in $(V - I)$ were observed. The main extinction coefficients in $B$, $V$, $R$ and $I$ as well as the color terms were calculated according to the following equations
\citep{lar99}:

$$U-u_0 = \alpha_U + \beta_U (u-b)_0$$
$$B-b_0 = \alpha_B + \beta_B (b-v)_0$$
$$V-v_0 = \alpha_V + \beta_V (b-v)_0$$
$$R-r_0 = \alpha_R + \beta_R (v-r)_0$$
$$I-i_0 = \alpha_I + \beta_I (v-r)_0$$

\noindent
where $B$, $V$, $R$ and $I$ will be the estimated magnitudes, $b$, $v$, $r$ and $i$ are the instrumental (and airmass-corrected) magnitudes. $\alpha$ and $\beta$ are the transformation coefficients for each filter.
\indent

    In a first iteration, a constant value associated with the sky background was subtracted from the images by using an interactive procedure that allows the user to select regions on the frame free of galaxies and bright stars.  Notice that uncertainities in determining the sky background may be the dominant source of uncertainities in the estimation of the magnitudes, color and surface brightness profiles.
\indent

    The most energetic cosmic-ray events were automatically masked using the IRAF task {\it cosmicrays} and field stars outside the bright visible disk of NGC 3367 were removed using the IRAF task {\it imedit} when necessary. Within the galaxy itself, care was taken to identify superposed stars. A final step in the basic reduction involved registration of all available frames for each galaxy and in each filter to within $\pm 0.1$ pixel. This step was performed by measuring centroids for foreground stars on the images and then performing geometric transformations using IRAF tasks {\it geomap} and {\it geotran}.

	All final images were convolved with a two dimensional gaussian function in such a way as to have a similar final resolution equivalent to the largest one given by the $V$ filter seeing, that is, full width at half maximum, fwhm$\sim2''.2$. 

\subsection{Uncertainities in the Photometry}

\subsubsection{As a result of the Method}
\indent
    An estimation of the uncertainities in our photometry involves two parts: 1) The procedures to obtain instrumental magnitudes and 2) the uncertainty when such instrumental magnitudes are transformed to a standard system. For 1), notice that the magnitudes produced at the output of the IRAF routines {\it qphot} and {\it phot} differ from each other as a result of different methods used in those procedures. Since we also have applied
extinction corrections to the instrumental magnitudes in this step, our estimation of the uncertainities are mainly concerned with these corrections and the estimation of the airmass. After a least square fitting, the associated uncertainities to the slope for each principal extinction coefficient are; $\delta(k_{U}) \sim 0.038$, $\delta(k_{B})
\sim 0.038$, $\delta(k_{V}) \sim 0.035$, $\delta(k_{R}) \sim 0.025$ and $\delta(k_{I}) \sim 0.025$. An additional uncertainity $\delta(airmass) \sim 0.005$ from the airmass routines in IRAF was also considered.
\indent

    For 2), the zero point and first order color terms are the most important to consider. After transforming to the standard system, by adopting our best-fit coefficients, the formal uncertainities from the assumed relations for $\alpha$  were 0.05, 0.05, 0.04, 0.04 and 0.04 in $U$, $B$, $V$, $R$ and $I$ and 0.04, 0.03, 0.03 and 0.04 for $\beta$. To estimate the total uncertainity in each band, it is necessary to use the transformation equations and then propagate the errors. Total typical uncertainties are 0.1, 0.1, 0.1, 0.09 and 0.1 in $U$, $B$, $V$, $R$ and $I$ bands, respectively.
\indent

    The estimated total magnitudes in this work were compared against other estimations
reported in the literature. This has been done for the standard stars and for NGC 3367 in common with other works.

\subsubsection{Standard Stars as calibrators}

    Our estimated CCD magnitudes for the nine stars in M67 are listed in Table 3 along with magnitudes and cross reference identification numbers reported in the literature \citep{gui91,che91}. A plot of our CCD estimated magnitudes versus those reported by Guilliland et al., (1991) and Chevalier \& Ilovaisky (1991) are shown in Figure 1
with no significant deviations, a continuous line with slope 1 is also plotted. These results suggest, a $\sigma \sim 0.12$, as the typical magnitude uncertainity. This is in agreement with our prior uncertainity estimations.
\clearpage
\newpage
\begin{figure}[tbh]
\includegraphics[width=10cm,height=10cm]{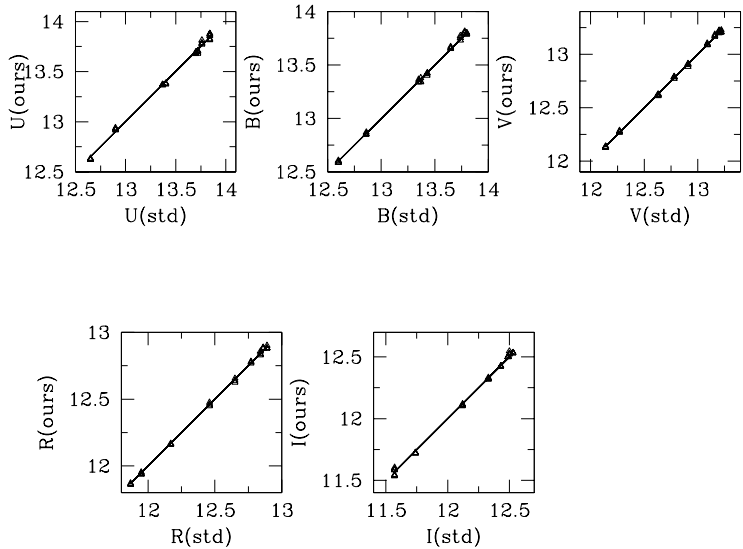}
\figcaption{Comparison between our estimated magnitudes
and those from Gilliland et al. (1991) [$U$, $B$, $V$ and $R$ bands] and Chevalier \& Ilovaisky (1991) [$I$ band] for 9 stars in common taken as standards \label{fig. 1}}
\end{figure}

\section {Results}

\subsection{Total Magnitudes and Colors}

    In this study, we have estimated total magnitudes for $U, B, V, R, I$ computed from different circular apertures chosen interactively to assure that they are large enough to contain the whole galaxy and still small enough to limit the uncertainities due to bad estimation of the flux from the background and any field stars. Finally we chose 
an aperture with a diameter of 140$''$ equivalent to 2$'.33$; this is slightly larger 
than the 2$'.02$ major diameter as given in NED. There is, however, a foreground star within the disk of NGC 3367 which we did not remove before estimating the global magnitudes and colors for NGC 3367. The star is object 82 in Tables 7 and 9 and it lies about 29$''$ to the 
north-west of the nucleus; we did not attempt to remove it because it is surrounded with emission from stellar structures in the disk and thus it makes it difficult to estimate the intensity level to replace the star with; star ID 82 turns out to be a weak star with magnitudes $U\sim17.9~B\sim17.9~V\sim17.4~R\sim16.9~I\sim16.2$. Total magnitudes from NGC 3367 were estimated in each band by using the {\it phot} routines in IRAF. Our final estimated  magnitudes for NGC 3367 were corrected for Galactic extinction (but not for intrinsic extinction); they are $U=11.70\pm0.1$, $B=11.91\pm0.1$, $V=11.52\pm0.1$, $R=11.05\pm0.1$, and $I=10.43\pm0.1$. The colors of NGC 3367 are $U - B = -0.21,~B - V = 0.39,~B - I$ = 1.48. 

\clearpage
\newpage
\begin{figure}[tbh]
\includegraphics[width=10cm,height=10cm,angle=-90]{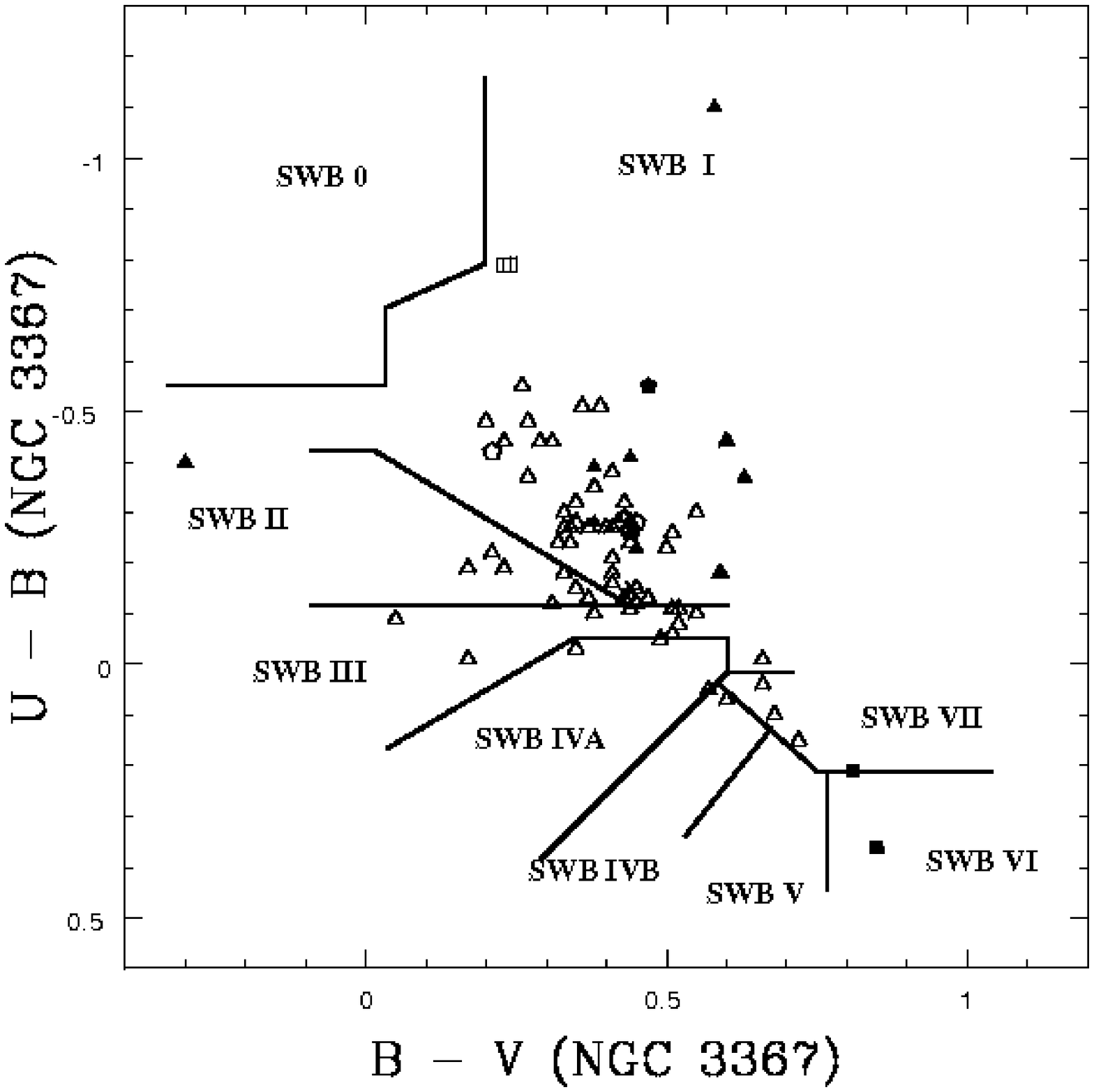}
\figcaption{The color color diagram of different structures throughout NGC 3367 is shown. Filled hexagon is the compact nucleus (ID 1), open squares are the bright star associations embedded in HII regions to the west of the on the bright semicircle rim of the disk (ID 2 and 3), open hexagons are the bright stellar associations embedded in HII regions east of the end of the bar (ID 4 - 6), filled squares are stellar associations to the northeast and southwest of the bar (ID 37 and 38 respectively), asterisk is the region which includes supernovae 2003aa (ID 42), open triangles are stellar associations in the disk, filled triangles are stellar associations beyond the semicircle bright end of the western side of the disk (ID 52 and 53). SWB type 
\citep{sea80,bic96}, SWB 0 have estimated ages between 0 and $10^7$ yrs, SWB I between $10^7$ and $3\times10^7$ yrs, SWB II between $3\times10^7$ and $7\times10^7$ yrs, SWB III between $7\times10^7$ and $2\times10^8$ yrs, SWB IVA between $2\times10^8$ and $4\times10^8$ yrs, SWB IVB between $4\times10^8$ and $8\times10^8$ yrs, SWB V between $8\times10^8$ and $2\times10^9$ yrs, SWB VI between $2\times10^9$ and $5\times10^9$ yrs, and SWB VII between $5\times10^9$ and $1.6\times10^{10}$ yrs. 
\label{fig. 2}}
\label{color color diagrams}
\end{figure}

\indent

    Since the aperture to estimate the magnitudes for the stellar structures throughout NGC 3367 must be small in order to include only star associations or small structures, it was decided to use an aperture with radius 8 pixels equivalent to $3''.41\sim 715$ pc and use IRAF task {\it phot}. One compared the estimated magnitudes, with an aperture of 8 pixels, for the bright stars in the field south west of NGC 3367 with those listed in Table 5 (with aperture radius of 26 pixels). The magnitude differences vary according to the star and the filter. For star ID 86 the difference in empirical magnitudes are $\Delta U(8 - 26) = 0.07$, $\Delta B(8 - 26) = 0.09$,
$\Delta V(8 - 26) = 0.10$, $\Delta R(8 - 26) = 0.09$, and $\Delta I(8 - 26) = 0.17$; while for star ID 91 the differences in magnitudes are $\Delta U(8 - 26) = 0.06$, $\Delta B(8 - 26) = 0.08$, $\Delta V(8 - 26) = 0.11$, $\Delta R(8 - 26) = 0.10$, and $\Delta I(8 - 26) = 0.18$, fair is to say that the magnitudes with aperture radius of 8 pixels have been corrected for Galactic extinction. Magnitudes differences for each filter for both stars are similar indicating indeed that using a smaller aperture radius gives an artificial weaker object by the amount given in each filter but the differences are within the values for the uncertainities of our method. The estimated magnitudes for the different structures throughout NGC 3367 (stellar associations) can be taken as reliable and are listed in Table 6. Table 7 lists the estimated magnitudes for the bright stars and a background galaxy in the field of NGC 3367. Table 8 lists the different colors $U - B$, $B - V$ and $V - I$ for the stellar structures throughout NGC 3367, while table 9 lists the colors of the stars and the small galaxy in the field of NGC 3367. See Tables 6 for comments on individual objects.

    Our estimated magnitudes for the whole galaxy in $U$, $B$, $V$ and $I$ bands and those reported in the RC3 Catalogue \citep{vau93} are in good agreement. The Galactic extinction towards NGC 3367 in the different filters are: A$_U=0.15$, A$_B=0.12$, A$_V=0.09$, A$_R=0.08$, and A$_I=0.05$ \citep{sch98}.

\indent

\subsection{$U B V R I$ Photometry of Star Associations and structures}

        It is a complicated problem to carry out photometry of stellar structures located within the disk of a galaxy, partly because of the strongly varying background, and partly because the structures are not perfect point sources (in other words, there must be a varying filling factor) nor circles (as it was our aperture). For disk galaxies, the colors $U - B$, $B - V$, $V - I$ are closely related to spectral type and show that there exist remarkable differences in stellar content \citep{sea73}. The colors of structures in a given galaxy are most certainly affected by internal reddening and chemical enrichment. In a particular galaxy, as in our study, the colors of different structures would indicate the contributions of the component stars in star clusters, very young star clusters embedded in HII region or compact HII regions, stellar associations without gas, associations embedded in HII regions and star clusters and associations with traces of emission [e.g. \citep{bic96}]. 

	It was decided in our photometric study of structures in NGC 3367 to estimate the color magnitudes of different structures using IRAF task {\it phot} with an aperture with radius 8 pixels (3$''.4$) which translates, in NGC 3367, to a radius with linear scale of 715 pc. No aperture correction were applied to our estimated magnitudes of the different structures. The photometry of all structures was obtained with the IRAF task {\it phot} which takes an area of the sky, well off the extention of the disk of the galaxy in study, as a background to be compared with.	

	In order to have an estimation of how good were our derived magnitudes for the stellar structures throughout NGC 3367, we first estimated the magnitudes of bright field stars to the south west of NGC 3367 (objets ID's 85 and 90 in Table 7) utilizing two IRAF tasks, {\it qphot}, and {\it phot}; each of these stars is relatively isolated and thus we could use apertures with different radii in order to estimate their magnitudes in the five filters. Table 4 gives our estimated magnitudes for the different filters for two bright stars in the south west of the field of NGC 3367 utilizing IRAF task {\it qphot} using an aperture radius of 23 pixels, equivalent to $9''.80$, and a width of ring (in order to estimate the brightness of the sky) of pixels, equivalent to width of ring of $2''.13$. Additionally IRAF task {\it phot} was used with three different circular apertures, one with a radius of 20 pixels, equivalent to $8''.52$, other with radius 23 pixels, equivalent to $9''.80$ and another one with 26 pixels, equivalent to $11''.08$. The estimated magnitudes for the five filters, $U$, $B$, $V$, $R$, and $I$, are listed in Table 5. The largest uncertainity observed in Table 5 using three different apertures amounts to $\sim0.05$ mag in $U$, $B$, $V$, and $R$ and it is $\sim0.13$ mag in $I$ for both stars. Comparison between estimated magnitudes using IRAF tasks {\it qphot} and {\it phot} for these field stars may be estimated comparing the results listed in Tables 4 and 5. The difference amounts in the worst case to 0.12 mag indicating that indeed IRAF task {\it phot} obtains a brighter value in each filter with the largest aperture; however the difference in magnitudes using different apertures with IRAF task {\it phot} is at worst $\sim0.03$ in the $I$ filter between using an aperture's radius of 20 pixels and an aperture's radius of 26 pixels. 

\indent

    The structures were chosen from the images in the $I$ and $U$ filters by eye at several locations of the disk of the galaxy with emphasis in choosing regions at different positions within NGC 3367, for example: regions at larger distances than the bright south-west semicircle (more than 10 kpc from the center, easily observed in images from the Palomar Sky Surveys), from the stellar bar's ends, from structures along and on spiral arms, from structures just beyond the stellar bar, etc. in order to have enough of them as to be able to compare their colors and estimate their ages. It included structures on the outside of the intense semicircle structure with bright surface brightness to the west of the disk. 

\indent
    Table 6 lists the positions, $\alpha$(J2000.0) and $\delta$(J2000.0) $\footnote{The peak of the $I$ image was anchored (in spatial position) to the peak of the high resolution radio continuum emission at 8.4 GHz \citep{gar02} to the position $\alpha(J2000)=10^h\;46^m\;34^s.956 \;\; \delta(J2000)= +13^{\circ}\;45'\;02''.94$. Although this is a logical supposition, it needs to be verified since in other galaxies the peak of the radio continuum emission does not coincide with the compact optical nucleus as a result of high extinction.}$ of the different structures
throughout NGC 3367, where we have estimated their magnitudes in the $U, B, V, R$ and $I$ broadband filters {\it corrected} for Galactic extinction. Table 8 is similar to Table 6 and lists the colors $U - B$, $B - V$, and $V - I$ corrected for Galactic extinction of each stellar structure in NGC 3367 along with their estimated SWB type.
\indent

    We have estimated the colors of the area in front of the nucleus; the estimated magnitudes were not corrected for intrinsic extinction; a visual extinction has been 
estimated, from blue spectroscopy, to be only A$_V\sim0.9$ \cite{cid05}. True magnitudes must be brighter since it is known that there is a lot of molecular gas in the central region of NGC 3367 \citep{gar05}.  At the distance of NGC 3367 (43.6 Mpc) the absolute magnitude is M$_V$ = -13.24 taking into account the Galactic optical extinction. As a comparison, the compact nucleus in the barred spiral galaxy NGC 4314, a LINER with a circumnuclear structure \citep{gar91}, has the following magnitude and colors within an aperture of 6$''$, 290 pc: $V = 14.07$, $U - B = 0.51$, and $B - V = 1.11$ \citep{ben80}, M$_V$ = -10.93 (at 10 Mpc taking into account the galactic optical extinction); the compact nucleus of NGC 4314 is reported to have a blue magnitude of $B = 16.6$ mag per square arc second \citep{lyn73}.

\indent

    Figure 2 shows the $U - B$ versus $B - V$ color color diagram for the different structures throughout the disk of NGC 3367. As a comparison, Figure 3, shows the corresponding $U - B$ versus $B - V$ diagram for the Large Magellanic Cloud \citep{bic96} stellar associations embedded in HII regions (NA in the nomenclature of Bica et al.), star clusters (C), very young clusters embedded in HII region or compact HII region (NC), stellar associations without gas (A) and a supernovae remanent. Previuos $U, B, V$ photometry of the SMC and LMC star clusters indicated that young clusters in the SMC were concentrated along its bar, while old clusters showed no concentration to the bar \citep{van81}; while the young clusters in the LMC were in and around the bar \citep{van81}. The difference between the color color diagrams between the clusters in the magellanic clouds and the structures in NGC 3367 is noteworthy: the LMC has many very young clusters with colors equivalent to the SWB 0 type \citep{sea80,bic92,bic96} $\footnote{The location of star clusters of different ages, in the $U - B$ versus $B - V$ color diagram delineate an average sequence known as SWB 0 to SWB VII \citep{sea80}. }$ (see section 
on ages below) and might also be the result of filling factor (radius of the aperture used, since the LMC is much closer than NGC 3367), intrinsic extinction and different chemical enrichment in NGC 3367.

\clearpage
\newpage

\begin{figure}[h]
\includegraphics[width=10cm,height=10cm,angle=-90]{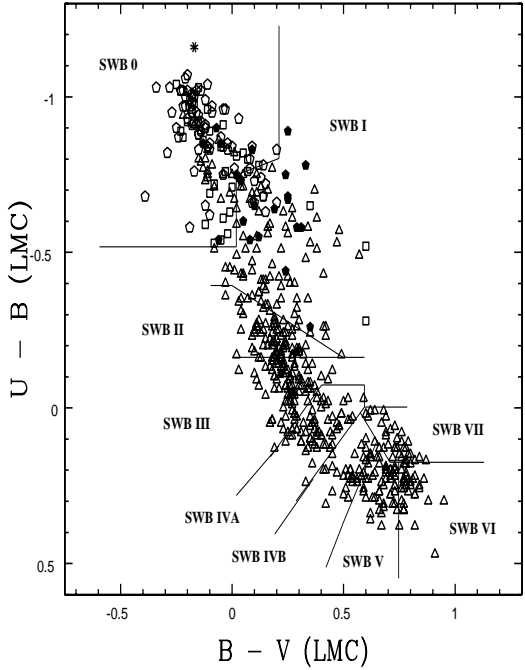}
\figcaption{The color color diagram of different stellar associations in the Large Magellanic Cloud \citep{bic96} is shown as a comparison. In their nomenclature, open polygons are stellar associations embedded
in HII regions (NA, $U - B$ in the interval from -1.1 to -0.55 and $B - V$ in the interval -0.4 to 0.2), filled polygons are stellar associations without gas (A, $U - B$ in the interval from -0.9 to -0.5 and $B - V$ in the interval -0.1 to 0.4), open squares are very young clusters embedded in HII regions or compact HII regions (NC, $U - B$ in the interval from -1.05 to -0.3 and $B - V$ in the interval -0.3 to 0.65), and open triangles are star clusters (C, $U - B$ in the interval from -0.95 to +0.5 and $B - V$ in the interval -0.25 to 0.95) \citep{bic96}.
\label{fig. 3}}
\end{figure}

\subsection{{\it U B V R I} Images}

    This section shows various images of NGC 3367. Figure 4 is the image in the $U$ filter in a grey scale and contours, proportional to surface brightness, showing the different structures throughout NGC 3367.

\clearpage
\newpage

\begin{figure}[t]
\includegraphics[width=12cm,angle=-90]{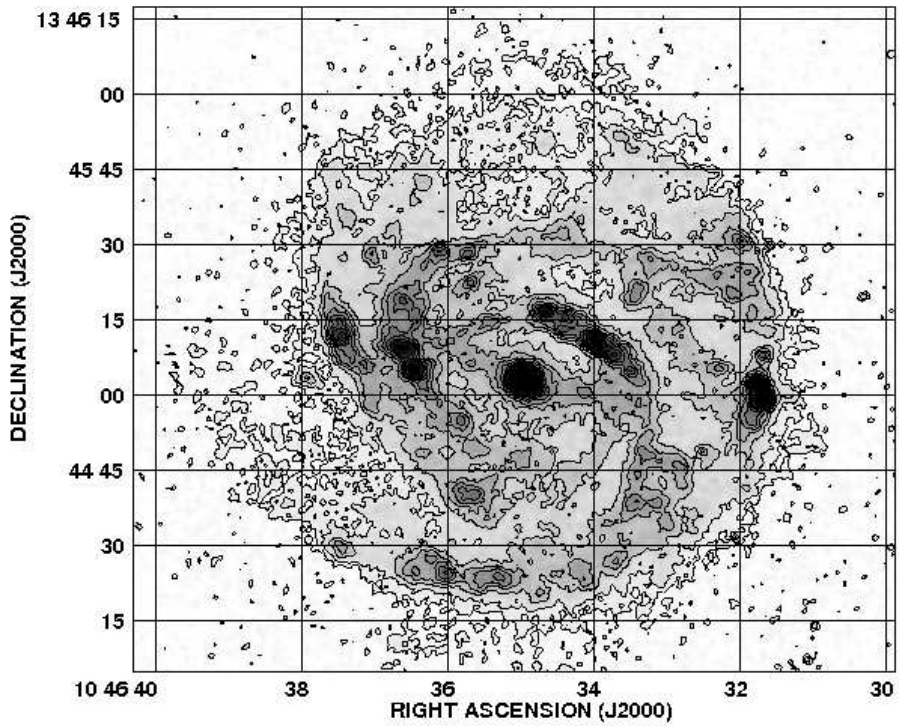}
\figcaption[figure=garcia-ubvri-fig5.ps]{Optical image, of NGC 3367, is shown in the broadband filter in the Ultraviolet in greyscale and contours. Contours are in mag arcsec$^{-2}$, and approximately correspond to m$_U\sim23.6$, 22.6, 21.8, 21.4, 21.1, 20.7, 20.4, 19.6, 18.8, 
18.1, and 17.4.
\label{fig. 4}}
\end{figure}

\clearpage
\newpage

\begin{figure}[tbh]
\includegraphics[width=12cm,height=13cm]{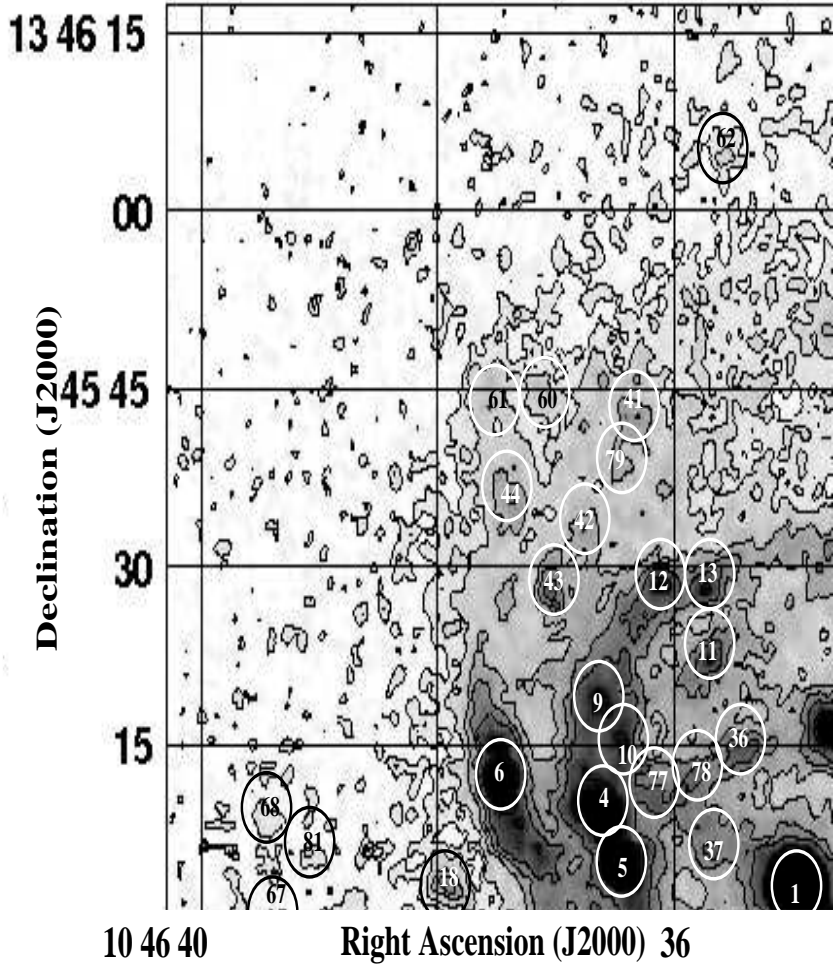}
\figcaption[figure=f5v3.eps]{North east quadrant of the $U$ image showing with circles the positions of different stellar structures; the circle is $\sim6''.8$ in diameter which, at the distance of NGC 3367, corresponds to a linear scale of 1.430 kpc.
\label{fig. 5} }
\end{figure}

\clearpage
\newpage

\begin{figure}[tbh]
\includegraphics[width=12cm,height=13cm]{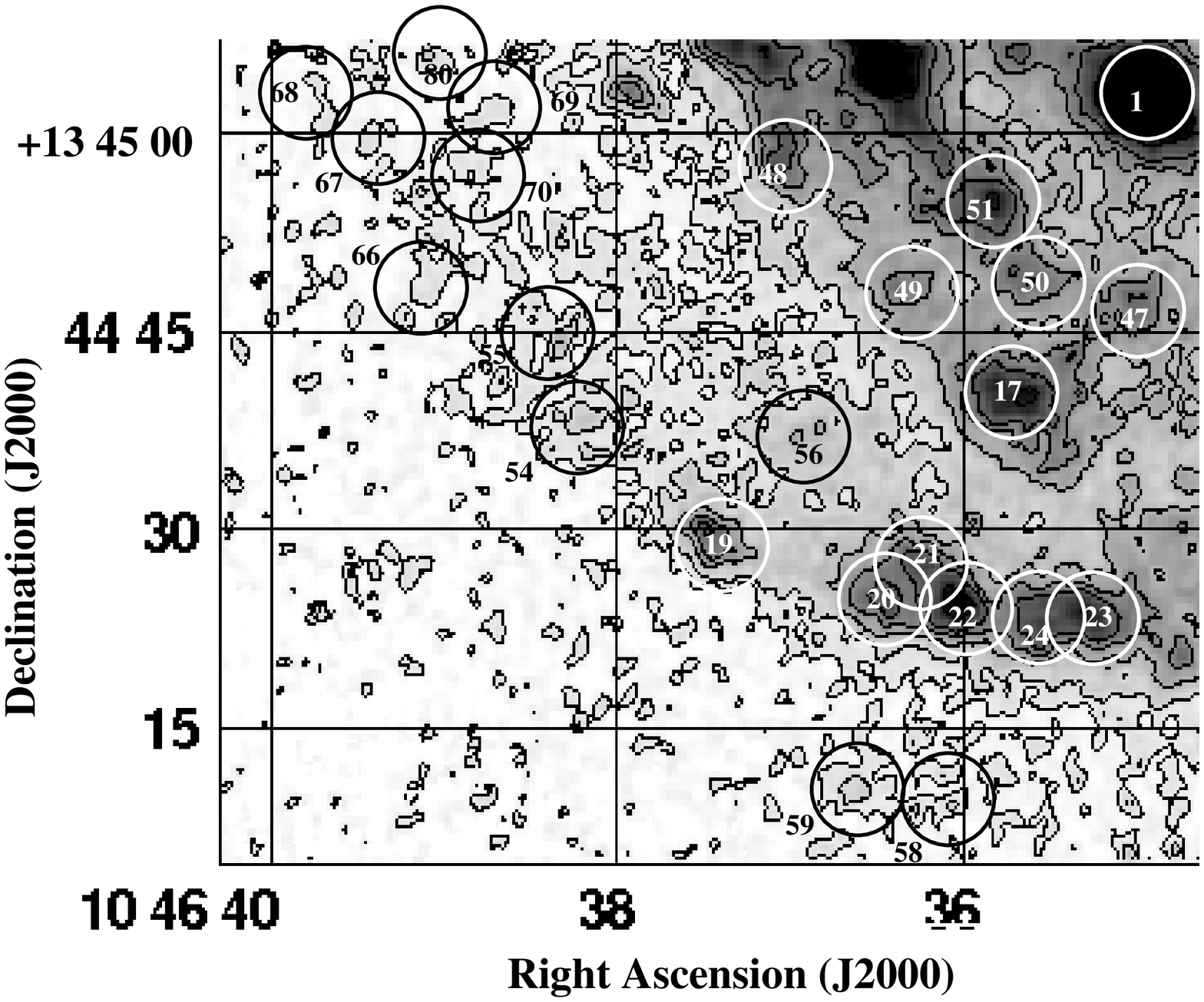}
\figcaption[figure=garcia-ubvri-f6.ps]{As for Fig. 5 but showing the Southeast quadrant.\label{fig. 6} }
\end{figure}

\clearpage
\newpage

\begin{figure}[tbh]
\includegraphics[width=12cm,height=13cm]{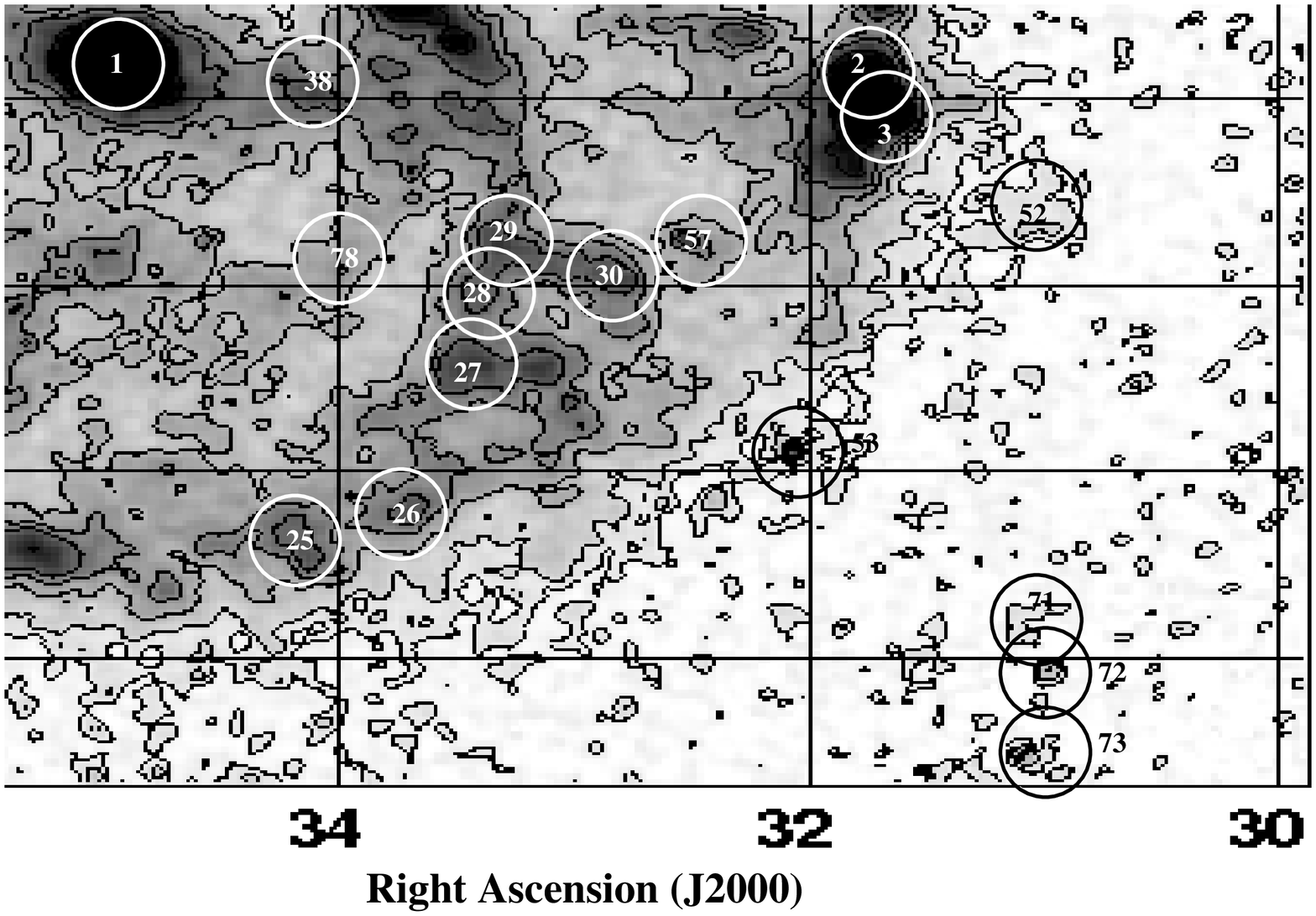}
\figcaption[figure=garcia-ubvri-fig7.ps]{As for Fig. 5 but showing the Southwest quadrant.\label{fig. 7} }
\end{figure}

\clearpage
\newpage

\begin{figure}[tbh]
\includegraphics[width=12cm,height=13cm]{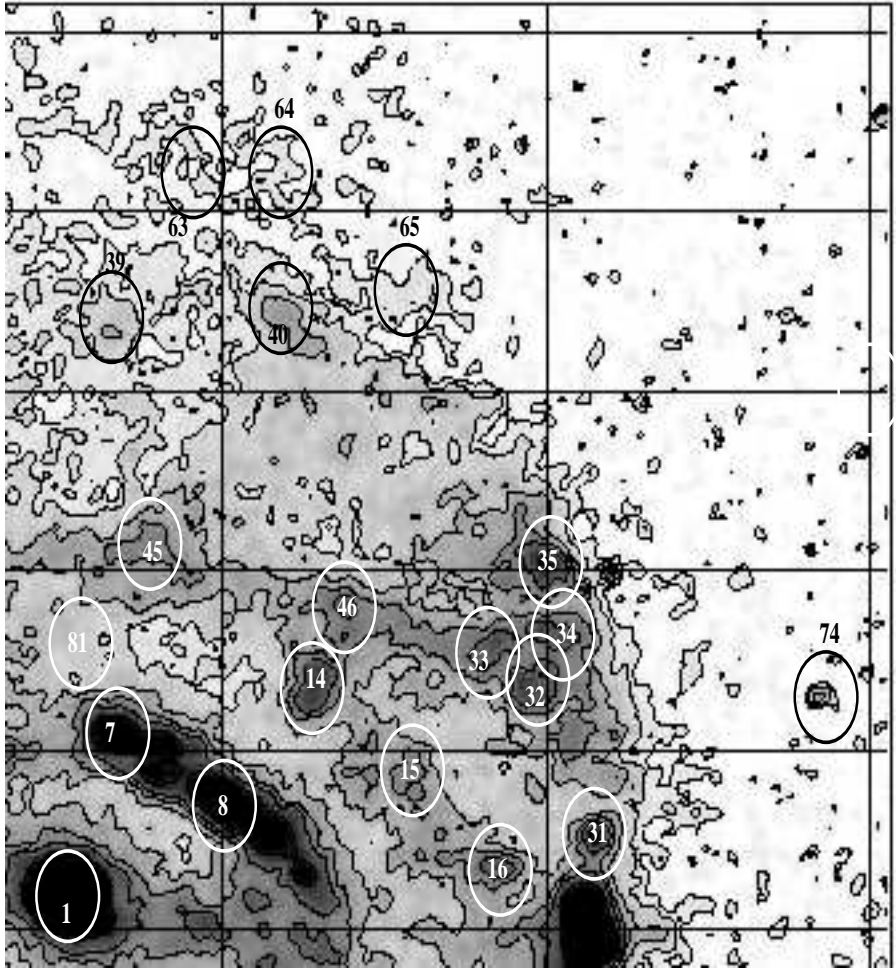}
\figcaption[figure=garcia-ubvri-fig8.ps]{As for Fig. 5 but showing the Northwest quadrant. \label{fig. 8}}
\end{figure}

\clearpage
\newpage

    Figures 5, 6, 7, and 8 show the four different quadrants of NGC 3367 with a circle indicating the position of the structure observed and its number, according to table 6, with the circle having the diameter of the aperture utilized for the photometry, that is $\sim6''.8$ which at the distance of NGC 3367 translates into a linear scale of $\sim1.4$ kpc. Figure 9 shows the $I$ image of the galaxy with a larger area in order to indicate the field stars and the 
background spiral galaxy. Figures 10, 11 and 12 show the $U - B$, $B - V$ and $V - I$ color images of NGC 3367. 

\begin{figure}[h]
\includegraphics[height=10cm,width=10cm]{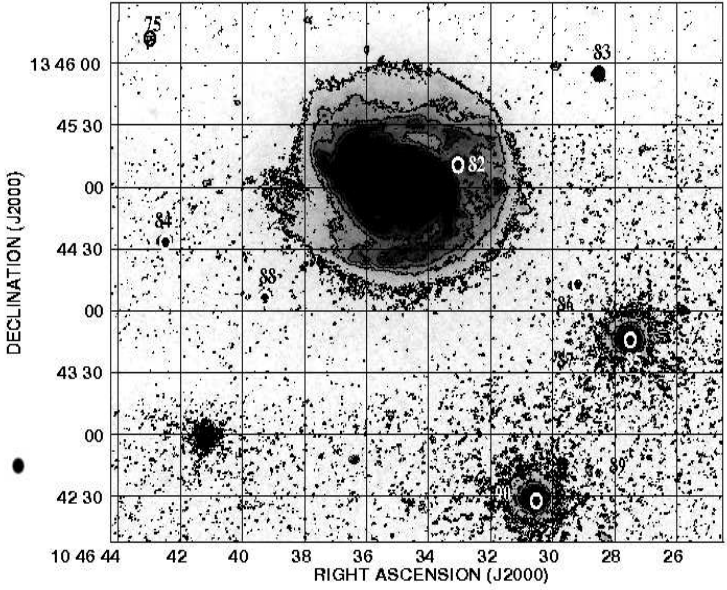}
\figcaption[figure=fig9v3.ps]{Optical image of NGC 3367 in the broadband $I$ filter is shown in greyscale and contours; contours are in mag arcsec$^{-2}$ and approximately correspond to 
m$_\sim20.2$, 19.2, 18.7, 18.4, 18.2, and 18. Circles indicate the positions of 
the stars in the field of NGC 3367, except for circle in the upper left corner that 
indicates the position of stellar structure ID 75; each circle is 8 pixels in radius 
or $\sim3''.4$. Stars 86 and 91 were used to estimate the uncertainities in our 
photometry when using IRAF tasks {\it qphot} (with one aperture) and {\it phot} with 
three different circular apertures. ID 90 is a background (disk) galaxy. 
\label{fig. 9}}
\end{figure}
\begin{figure}[tbh]
\includegraphics[width=10cm,height=10cm,angle=-90]{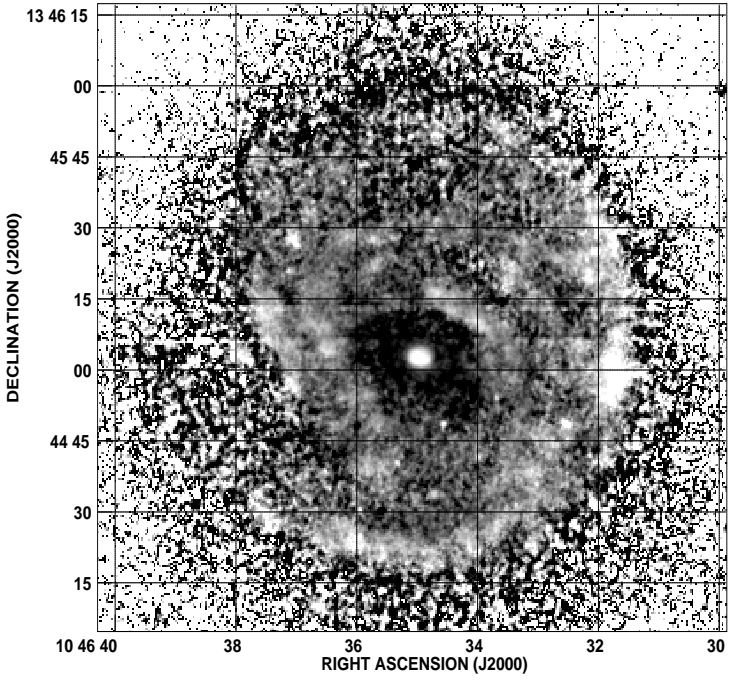}
\figcaption{$U - B$ color image of the disk galaxy NGC 3367: Greyscale is in arbitrary units in such a way that white areas indicate brighter ultraviolet emission. \label{fig. 10}}
\end{figure}

\clearpage
\newpage

\begin{figure}[tbh]
\includegraphics[width=10cm,height=10cm,angle=-90]{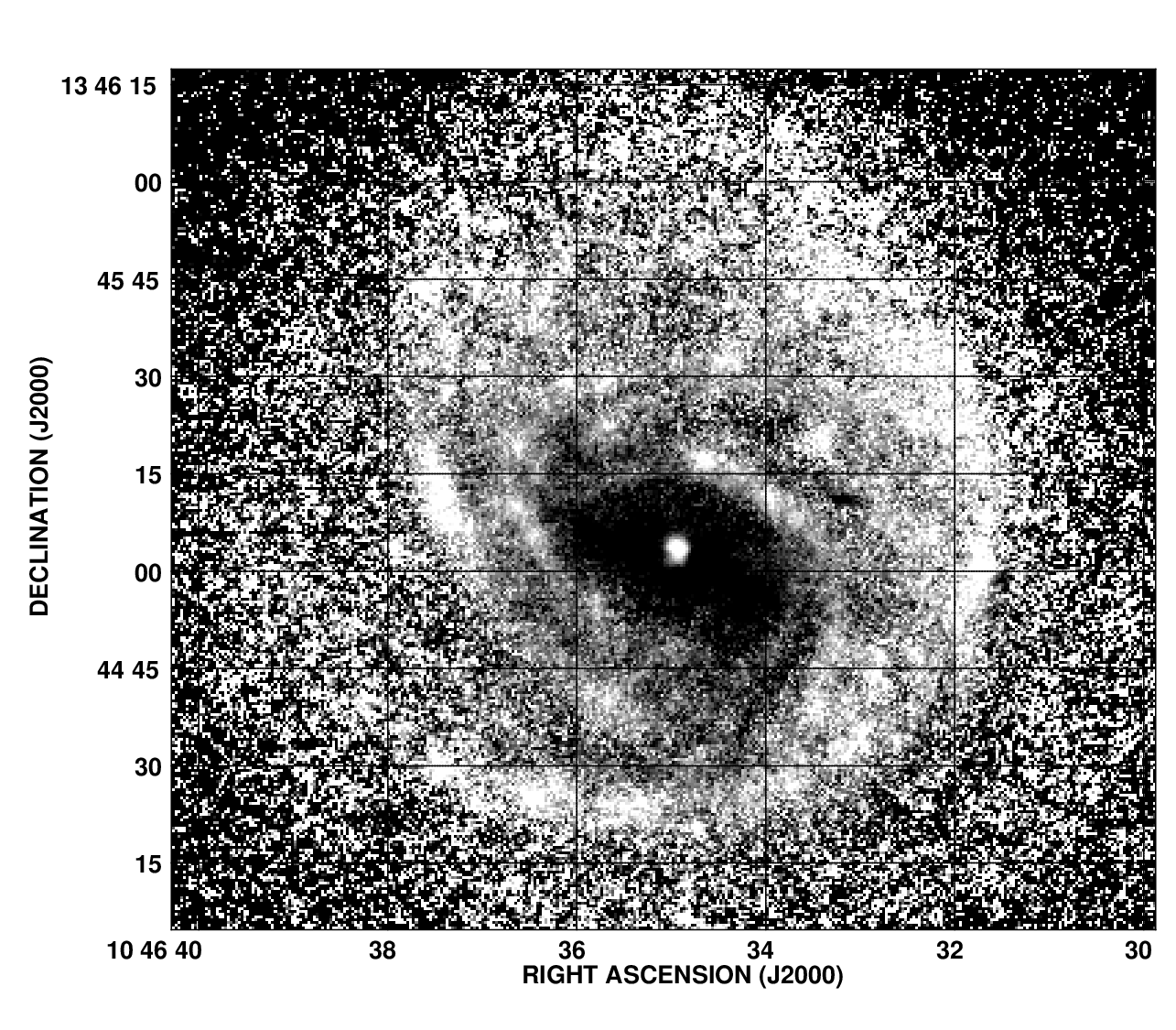}
\figcaption{$B - V$ color image of the disk galaxy NGC 3367. \label{fig. 11}}
\end{figure}

\clearpage
\newpage

\begin{figure}[tbh]
\includegraphics[width=10cm,height=10cm,angle=-90]{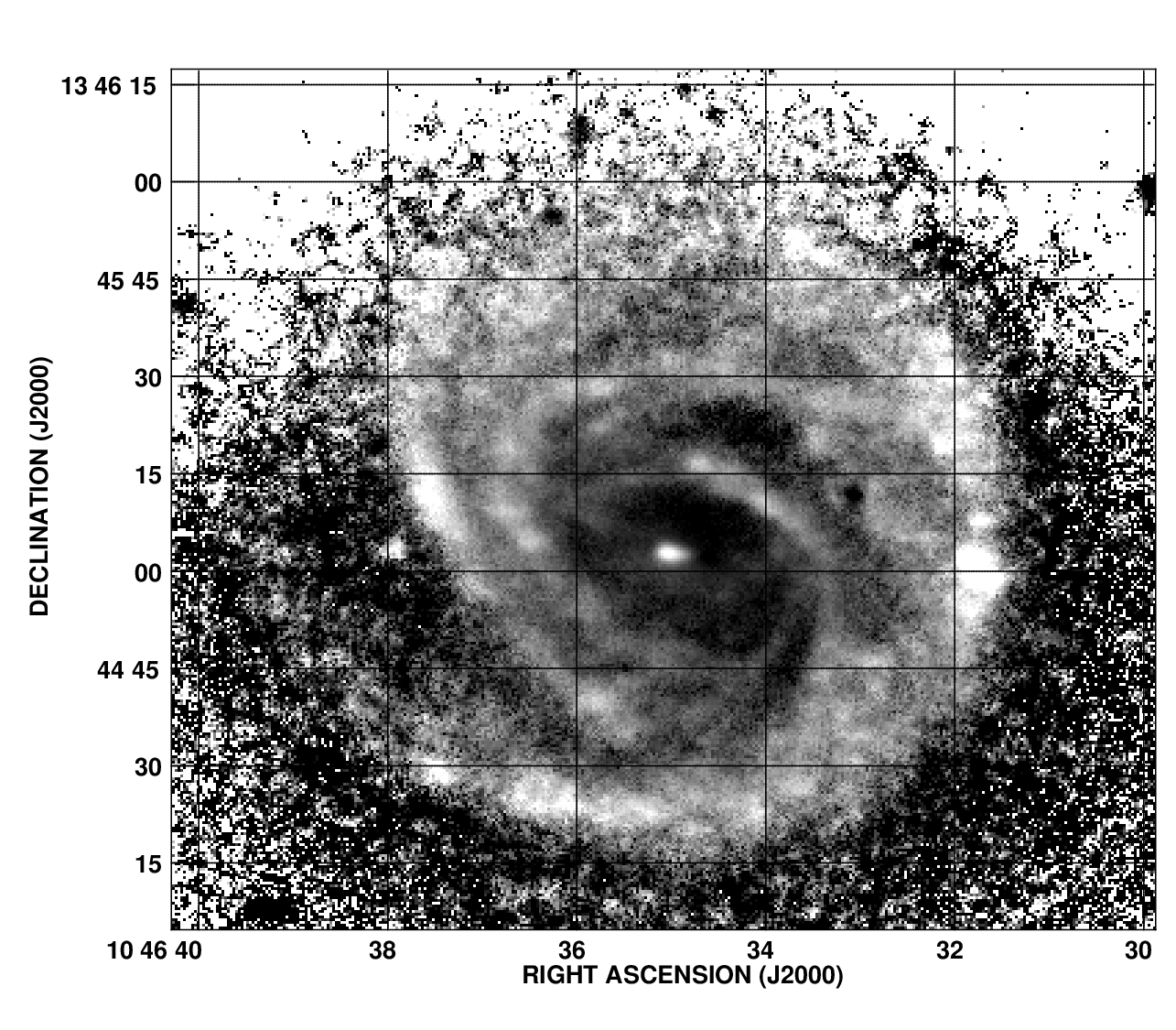}
\figcaption{$V - I$ color image of the disk galaxy NGC 3367. \label{fig. 12}}
\end{figure}

\section{Discussion}

\subsection{Spatial distribution of structures with different SWB type}
\indent

      There are 42 stellar structures with age type SWB I $\footnote{Approximate intervals of ages to different SWB classes are as follows \citep{sea80,bic96}: SWB 0 have ages between 0 and $10^7$ yrs, SWB I have ages between $10^7$ and $3\times10^7$ yrs, SWB II have agesbetween $3\times10^7$ and $7\times10^7$ yrs, SWB III have ages between $7\times10^7$ and $2\times10^8$ yrs, SWB IVAhave ages between $2\times10^8$ and $4\times10^8$ yrs, SWB IVB have ages between $4\times10^8$ and $8\times10^8$ yrs,SWB V have ages between $8\times10^8$ and $2\times10^9$ yrs, SWB VI have ages between $2\times10^9$ and $5\times10^9$yrs, and SWB VII have ages between $5\times10^9$ and $1.6\times10^{10}$ yrs \citep{chi88,gir93,and03}. }$ out of 81 observed which correspond to 51\%. A natural question is, where is their spatial location? Structure ID 1 corresponds to the central innermost stellar association which includes the nucleus of the galaxy; ID 2 and 3 are two stellar associations embedded in bright HII regions on the western bright surface brightness edge of the disk at about 10 kpc from the center; ID 4 and 5 lie on an inner eastern bright surface brightness area which seems to be a spiral arm or part of an inner ring (just beyond the edge of the stellar bar) at about 4.5 kpc from the center; ID 6 and 48 lie on another structure on an eastern spiral arm, parallel to the arm where ID 4 and 5 lie, at about 8 kpc from the center; ID 7 and 8 lie on the inner most spiral arm north-west of the center; ID 14 lies on the crooked spiral arm north west of the center at about 6 kpc; ID 18 lies at about 10 kpc east of the center just outside of the second spiral arm; ID 19 - 22, 25, 27 all lie on the southern spiral arm; ID 30 lies on a circular surface brightness structure south-west of the south-west end of the stellar bar; ID 31 - 35 and 40  all lie on the western spiral arm; ID 43 lies on an interarm on the eastern region of the disk at about 8.5 kpc; ID 57 lies on an interarm region west of the nucleus at about 8 kpc.

	Structures ID's 52 - 55, 58, 59, 62 - 74, and 80 lie outside the bright surface brightness semicircle at distances more than about 13 kpc. The features are detected and shown in Figs. 4, 5, 6, 7, and 8.
\indent

     It is important to emphasize that in the central region (ID 1), the western sources (ID 4 - 6) and the eastern sources (ID 2, 3), there is a large velocity dispersion of ionized gas, detected through H$\alpha$ \citep{gar01} and in the molecular gas CO \citep{gar05}. 
\indent

\begin{figure}[t]
\includegraphics[width=10cm]{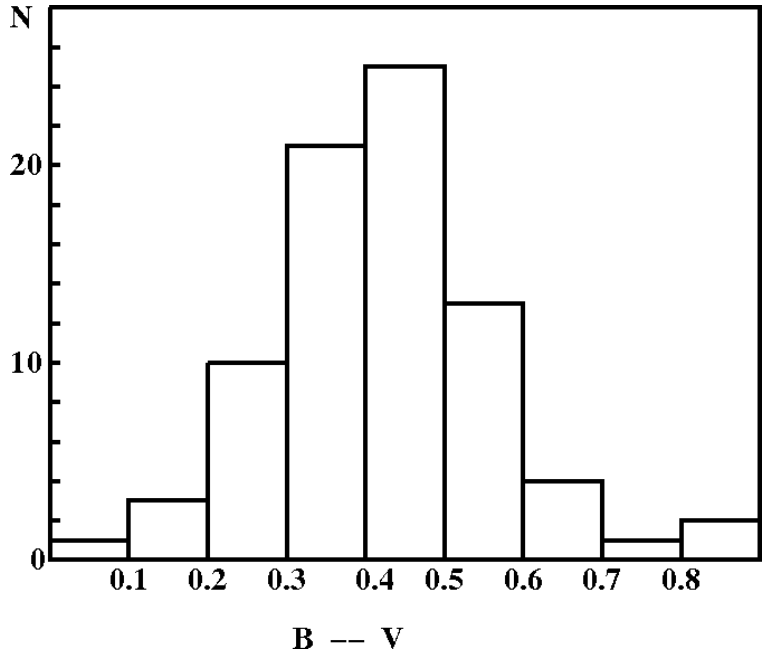}
\figcaption[figure=fig13v3.ps]{Histogram of number of structures studied in this work versus their $B - V$ color. Notice that 61 out of 81 of the structures have $B - V$ color less than 0.5 (blue). \label{fig. 13}}
\end{figure}

      In summary, most of the structures with age type SWB I lie close or on spiral arms. Perhaps the most important result of the spatial distribution of structures with age type SWB I is that several lie outside of the eastern bright surface brightness disk, ID 58 and 59 to the south, ID 65, 66, 69, 70, to the east while ID 53, 71 - 74 lie to the western side and ID 75 lies well outside the disk to the north-eastern end. The colors of structures ID 58, 59, 71 - 74 and 75 suggest that they are newly formed stellar structures and yet they lie at distances beyond the bright visible disk of NGC 3367, that is beyond 10 kpc from the center.

\subsection{Estimated ages of the structures throughout NGC 3367}

    It is difficult to speak about ages of the different structures throughout NGC 3367 mainly
because, as we have seen, different structures could indicate star associations embbeded in HII regions, very young star cluster embedded in HII regions, star associations not associated with gas, star associations in or around the nucleus, and other combinations. Among some of the problems with estimating the magnitudes of different structures are: filling factor, spatial distribution of structure within the aperture, and the fact that we used a circular aperture. We realize that with our circular aperture diameter, of $6''.8\sim1.4$ kpc, for estimating magnitudes and colors one can not detect individual star clusters in this galaxy. However, the study of star clusters with $U, B, V, R,$ and $I$ photometry and colors, and their derived ages in the LMC \citep{sea73,sea80} might be taken as an excellent reference to compare the $U, B, V, R,$ and $I$ colors of the different structures throughout NGC 3367 and {\it estimate} their ages if only one were to assume that the colors were equivalent, the chemical abundance were similar, have the same metallicity as in the LMC (which of course there is no reason to be an absolute valid assumption) and more important we might {\it use} the SWB ages if one ignores the internal reddening.  
\indent

     ID 42, which includes supernova 2003AA, (asterisk in Figs. 2 and 4) has colors similar to SWB II. Structures ID 52 and ID 53, beyond the bright western high surface brightness rim of NGC 3367 (filled triangles in Fig. 2) have equivalent colors to clusters SWB IVB and V. ID 37 and ID 38 are the structures north-east and south-west respectively of the stellar bar (filled squares in Fig. 2) have colors similar to SWB VI, indicating old stellar populations, but this result needs to be taken with caution since there the colors have not been corrected for intrinsic extinction as a result of the probable existence of molecular gas (CO, H$_2$) and dust 
at least to the southwest side of the bar \citep{gar05}.
\indent 

    Figure 12 shows the histogram of the number of structure throughout NGC 3367 versus the color $B - V$ where there is a large number of structures with blue colors, $B - V\leq0.5$. Figure 13 shows the histogram of age groups according to SWB types (assuming that the models for star clusters in the LMC also apply to the structures and angular resolution observed in NGC 3367) where 64\% of the structures observed have age types smaller or equal than SWB type II or less than 70 Myrs.

\begin{figure}[tbh]
\includegraphics[width=10cm,angle=-90]{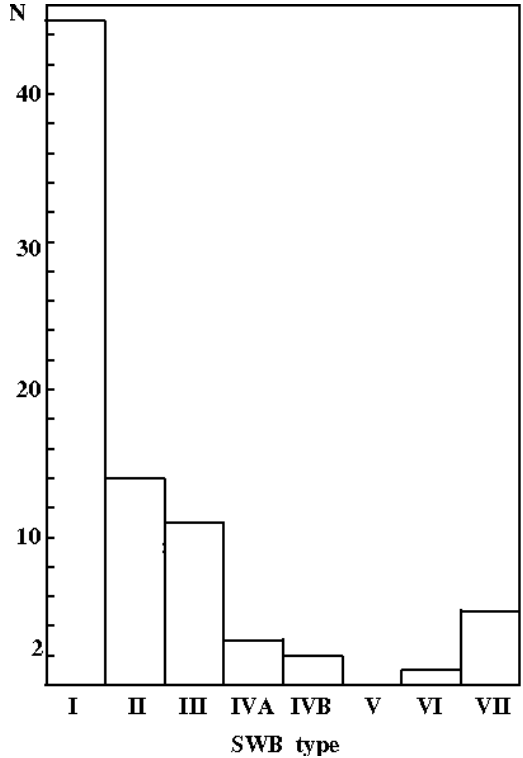}
\figcaption[figure=fig14v3.ps]{Histogram of number of structures studied in this work versus age group, SWB type. Notice that 86\% (70 out of 81) of the structures have SWB types I, II or III. This number would increase as a result of intrinsic extinction.\label{fig. 14} }
\end{figure}
Figs. 13 and 14 suggest that a large number of structures throughout NGC 3367 are relatively blue and young structures. This number of young structures would probably increase as one were able to somehow correct for intrinsic extinction, and different metallicity.

\section{Conclusions}

        To conclude, we have obtained several images in the broadband filters $U, B, V, R,$ and $I$ in order to estimate the optical magnitudes and colors of 81 different stellar structures throughout the disk of NGC~3367. The structures were chosen from the images in the $I$ and $U$ filters by eye at different locations of the disk of the galaxy in order to have enough of them as to be able to compare their colors as a function of their spatial location. It included structures to the north-east, south and west, well outside of the semicircle structure with bright surface brightness. 
\indent

	Our estimated magnitudes for the NGC 3367, after correcting for Galactic extinction are
$U=11.70\pm0.1$, $B=11.91\pm0.1$,$V=11.52\pm0.1$, $R=11.05\pm0.1$, and $I=10.43\pm0.1$. NGC 3367 global colors, after being corrected for Galactic extinction are $U - B = -0.21,~B - V = 0.39,~B - I$ = 1.48. All images except the image in $V$ were convolved with an appropiate two dimensional gaussian function in order to have a final of 2$''.2$ full width at half maximum, since the seeing was different for each filter and was about 1$''.7$ to $2''.2$. The 81 stellar structures reported in this paper were selected mainly from the $U$ and the $I$ images based on the location in the disk in order to have as many as possible from different positions in NGC 3367. We have made careful analysis of the calibration methods to estimate magnitudes and have compared different IRAF tasks, in particular, {\it qphot} and {\it phot} utilizing different apertures in order to estimate the uncertainities in the final magnitudes at different filters with 9 stars in M 67 as our standards.  In the case of estimating the magnitudes for the structures throughout the disk in NGC 3367, we have made careful analysis utilizing different IRAF tasks, in particular, {\it qphot} and {\it phot} using different aperture diameters for the bright stars in the field of NGC 3367 in order to estimate the uncertainities in the final magnitudes at different filters before estimating the final magnitudes reported for the different structures in NGC 3367. The magnitudes reported here for the structures in NGC 3367 were estimated using an aperture radius of 8 pixels ($\sim3''.4\sim715$ pc).
\indent

    We were able to estimate the optical magnitudes $U, B, V, R,$ and $I$ and colors $U - B,~B - V, U - V,$ and $V - I$ for 81 different stellar structures throughout the disk in NGC 3367. It is difficult to estimated the ages of the structures in NGC 3367 given our circular aperture diameter, equivalent to $6''.8\sim1.4$ kpc. However if their colors were taken as representative as the colors of star clusters (studied by many other authors with better angular resolution, appropiate chemical abundance and current theoretical star evolutionary tracks) in the LMC many have colors equivalent to cluster types SWB I, and SWB II suggesting that many are young associations with only tens of millions of years of age.
\indent

    We found 41 structures out of 81 (50\%) with age type SWB I. Several of those structures lie 
near or in spiral arms; some lie outside of the bright surface brightness visible disk (for example the POSS image).
\indent

        In order to understand better the large scale environment of NGC 3367 it would also be very important to determine the large scale spatial distribution and kinematics of atomic hydrogen (HI) and new observations with high angular resolution of the emission of the molecular gas (CO, H$_2$) in the center of NGC 3367.

\section*{Acknowledgements}

JAGB and HMHT thank the staff of the Observatorio Astronomico Nacional at San Pedro M\'artir, Baja
California, M\'exico for the help in the observations. HMHT acknowledges support from grant
CONACyT (M\'exico) 42810. JAGB acknowledges partial financial support from DGAPA (UNAM, M\'exico) grant IN107806-2. The authors would like to thank the anonimous referee for his (her) comments and suggestions on how to improve this paper.


\clearpage
\newpage

\begin{table}
\scriptsize{\caption[]{General Properties of the barred disk galaxy NGC 3367}
\noindent\begin{tabular}{lcr}
\hline
Characteristic     &            Value                   & Reference \cr
Hubble Type (RC3)  & SB(rs)c                            & 3 \cr
Nucleus spectral type & Sy 2-like, HII                  & 1,2 \cr
$U$ mag            & 11.89 $\pm$ 0.15                   & 3 \cr
$B$ mag            & 12.05 $\pm$ 0.14                   & 3 \cr
$V$ mag            & 11.50 $\pm$ 0.14                   & 3 \cr
$I$ mag            & 10.60 $\pm$ 0.07                   & 4 \cr
V$_{sys}$(HI)      & 3030  $\pm$ 8 km s$^{-1}$          & 5 \cr
V$_{sys}$(opt)     & 2850  $\pm$ 50 km s$^{-1}$         & 1 \cr
V$_{sys}$(FP$^a$)  & 3032 $\pm$ 3 km s$^{-1}$           & 6 \cr
V$_{sys}$(CO)      & 3035  $\pm$ 5 km s$^{-1}$          & 7 \cr
$\alpha$(J2000)    & 10$^h~46^m~34.95^s$               & 8 \cr
$\delta$(J2000)    & +13$^{\circ}~46'~02.9''$          & 8 \cr
log L$_B/L_{\odot}$ &  10.68                            & 9 \cr
Photometric major axis P.A. & 109$^{\circ}\pm 4$        & 10 \cr
Photometric inclination     &   6$^{\circ}\pm 1$        & 10 \cr
Major kinematic axis P.A. & 51$^{\circ}\pm 3$          & 6 \cr
Inclination of disk (from kinematics) & 30$^{\circ +2}_{~-5}$ & 6 \cr
Stellar bar P.A.     &  65$^{\circ}\pm 5$                & 11 \cr
Stellar bar diameter &  32$''\sim6.7$ kpc             & 11 \cr
M(HI)                & $7\times10^9$ M$_\odot$         & 5 \cr
M(H$_2$), r=27$''$   & $2.6\times10^9$ M$_\odot$       & 7 \cr
M(H$_2$), r=4$''$.5  & $5.9\times10^8$ M$_\odot$       & 7 \cr
Visual extinction, A$_v/2$ toward nucleus$^b$ & $\sim75$ mag & 7 \cr
Total Radio Continuum (20 cm) flux density & 119.5 mJy  & 12 \cr
Sum flux density (20 cm) from extended lobes & 39 mJy   & 8 \cr
Projected distance of each synchrotron lobe & $\sim$6 kpc & 8 \cr
Position angle of inner radio continuum synchroton emission & $\sim46^{\circ}\pm$ 6 & 8 \cr
Position angle of synchrotron lobes & $\sim30^{\circ}\pm 20$ & 8, 13 \cr 
\hline
\end{tabular}

References: 1) V\'eron-Cetty \& V\'eron (1986), 2) Ho, et al. (1997a), 3) de Vaucouleurs et al. (1993) RC3,
4) Nasa Extragalactic Database (NED), 5) Huchtmeier \& Seiradakis (1985), 6) Garc{\iac}a-Barreto \& Rosado (2001),
7) Garc{\iac}a-Barreto et al. (2005), 8) Garc{\iac}a-Barreto, Franco \& Rudnick (2002), 9) Tully (1988),
10) Grosb{\o}l (1985), 11) Garc{\iac}a-Barreto et al. (1996a,b), 12) Condon et al. (1998), 
13) Garc{\iac}a-Barreto et al. (1998)

$^a$ From Fabry-Perot H$\alpha$ optical interferometry; $^b$ Maloney (1990).}
\end{table}
\clearpage
\newpage

\begin{table}
\caption[]{Journal of observations, average seeing and exposure time.}
\noindent\begin{tabular}{cccccc}
\hline
            & $U$          & $B$          & $V$       & $R$          & $I$ \cr
Seeing      & $\sim2''.0$ & $\sim2''.1$ & $\sim2''.2$ & $\sim1''.80$ & $\sim1''.70$ \cr
$\Delta{\lambda}$ (\AA) & 600 & 950      & 1400      & 400          & 1600 \cr
$\lambda_{efective}^{central}$ (\AA) & 3540 & 4330 & 5750 & 6340     & 8040 \cr
M67         & 8$\times30^s$  & 8$\times30^s$ & 8$\times15^s$ & 8$\times5^s$  & 8$\times5^s$   \cr
NGC 3367    & 5$\times600^s$ & 5$\times600^s$ & 5$\times600^s$ & 8$\times300^s$ & 8$\times300^s$ \cr
\hline
\label{tabjournal}
\end{tabular}
\end{table}

\begin{table}
\scriptsize{ \caption[]{Our estimated UBVRI photometry of stars in M67 with IRAF {\it phot}, and 
those observed by other groups}
\begin{flushleft}
\noindent\begin{tabular}{cllllllllllll}
\hline
Our ID  & ID$^{1,2}$ & $U$(ours) & $U^1$ & $B$(ours) & $B^1$ & $V$(ours) & $V^1$ & $R$(ours)
& $R^1$ & $I$(ours) & $I^2$ \cr
\hline
1  & 48,14c          & 13.79 & 13.76 & 13.75 & 13.74 & 13.17 & 13.16 & 12.84 & 12.84 & 12.50 & 12.49\cr
2  & 16,15c          & 12.92 & 12.90 & 12.85 & 12.86 & 12.27 & 12.27 & 11.94 & 11.95 & 11.59 & 11.57\cr
3$^3$ & 49,17c       & 13.85 & 13.84 & 13.80 & 13.78 & 13.21 & 13.20 & 12.88 & 12.86 & 12.53 & 12.50\cr
4  & 28,18c          & 13.36 & 13.37 & 13.35 & 13.37 & 12.89 & 12.91 & 12.63 & 12.65 & 12.32 & 12.33\cr
5  & 52,19           & 13.84 & 13.84 & 13.79 & 13.80 & 13.21 & 13.22 & 12.88 & 12.89 & 12.53 & 12.53\cr
6  & 27,20c          & 13.37 & 13.40 & 13.35 & 13.35 & 12.78 & 12.78 & 12.46 & 12.46 & 12.10 & 12.12\cr
7  & 13,21c          & 12.62 & 12.65 & 12.59 & 12.60 & 12.13 & 12.14 & 11.86 & 11.87 & 11.54 & 11.57\cr
8  & 37,22c          & 13.68 & 13.70 & 13.41 & 13.43 & 12.61 & 12.63 & 12.16 & 12.17 & 11.72 & 11.74\cr
9  & 44,23           & 13.69 & 13.72 & 13.66 & 13.65 & 13.09 & 13.09 & 12.77 & 12.77 & 12.42 & 12.43\cr
\hline
\end{tabular}
\end{flushleft}
$^1$ \citep{gui91}; $^2$ \citep{che91}; $^3$ Position of star $\#3$ is: $\alpha$(B1950.0)=08$^h~48^m~37^s.5$,
$\delta$(B1950.0)=+11$^{\circ}~57'~23''.4$; as a comparison, the position and magnitude reported in the USNO-B Star
Catalog \citep{mon03}, for a star probably associated with star $\#3$ is: $\alpha$(B1950.0)=08$^h~48^m~37^s.95$,
$\delta$(B1950.0)=+11$^{\circ}~57'~57''.9$ mag$_{B1}=13.98$, mag$_{B2}=13.54$ }

\end{table}

\footnotesize
\begin{table}
\caption[]{UBVRI estimated photometry of bright stars in the field of NGC 3367 with IRAF task {\it qphot},
aperture (apr), and width of ring (wor)}
\begin{flushleft}
\noindent\begin{tabular}{lllccccc}
\hline
\multicolumn{8}{c}{apr = 23 pixels (9$''.8$),  wor = 5$''$}\\
\hline
ID   & $\alpha$(J2000.0)  & $\delta$(J2000.0)       &  U    &   B    &  V    &   R  & I \cr
\hline
85  & 10$^h~46^m~27^s.56$ & +13$^{\circ}~43'~46''.0$ & 13.91 & 13.51 & 12.87 & 12.47 & 12.22\cr
90  & 10$^h~46^m~30^s.60$ & +13$^{\circ}~42'~28''.6$ & 13.13 & 13.00 & 12.68 & 12.37 & 12.12\cr
\hline
\end{tabular}
\end{flushleft}
\end{table}

\footnotesize{
\begin{table}
\caption[]{UBVRI Comparative Photometry of bright stars in the field of NGC 3367, using IRAF task {\it phot}
with three aperture (apr) radii:
apr 20 pixels $\sim8''.5$, 23 pixels $\sim9''.8$ and 26 pixels $\sim11''.1$
}
\begin{flushleft}
\noindent\begin{tabular}{lllccccc}
\hline
\multicolumn{8}{c}{apr = 20 pixels ($8''.5$)}\\
\hline
ID     & $\alpha$(J2000.0)  & $\delta$(J2000.0)         &  U    &  B    &   V   &  R    & I \cr
\hline
85     & 10$^h~46^m~27^s.56$ & +13$^{\circ}~43'~46''.0$ & 13.87 & 13.46 & 12.82 & 12.43 & 12.12\cr
90     & 10$^h~46^m~30^s.60$ & +13$^{\circ}~42'~28''.6$ & 13.09 & 12.95 & 12.63 & 12.32 & 12.02\cr
\hline
\multicolumn{8}{c}{apr = 23 pixels ($9''.8$)}\\
\hline
85     & 10$^h~46^m~27^s.56$ & +13$^{\circ}~43'46''.0$ & 13.87 & 13.46 & 12.81 & 12.42 & 12.11\cr
90     & 10$^h~46^m~30^s.60$ & +13$^{\circ}~42'28''.6$ & 13.09 & 12.95 & 12.62 & 12.31 & 12.01\cr
\hline
\multicolumn{8}{c}{apr = 26 pixels ($11''.1$)}\\
\hline
85     & 10$^h~46^m~27^s.56$ & +13$^{\circ}~43'46''.0$ & 13.86 & 13.45 & 12.81 & 12.42 & 12.09\cr
90     & 10$^h~46^m~30^s.60$ & +13$^{\circ}~42'28''.6$ & 13.09 & 12.95 & 12.62 & 12.31 & 11.99\cr
\hline
\end{tabular}
\end{flushleft}
\end{table}
}
\clearpage
\newpage

\begin{deluxetable}{rrrrrrrr}
\tabletypesize{\tiny}
\tablecolumns{8} \tablewidth{0pc}
\tablecaption{$U B V R I$ Photometry, corrected
for Galactic extinction, of structures throughout NGC 3367 using 
IRAF task {\it phot} using an aperture with radius 8 pixels $\sim3''.41\sim715$ pc} 
\tablehead{ \colhead{ID} & \colhead{$\alpha$(J2000.0)} & \colhead{$\delta$(J2000.0)} &
\colhead{U} & \colhead{B} & \colhead{V} & \colhead{R} & \colhead{I}} 
\startdata 
1\tablenotemark{1} & 34$^s.949$\tablenotemark{2} & 45$'$ 02$''.8$\tablenotemark{2} & 14.84 & 15.39 & 14.92 & 14.39 & 13.92\\
2  & 31$^s$.71 & 45$'$ 00$''.0$ & 16.49 & 17.28 & 17.05 & 16.82 & 16.45\\
3  & 31$^s$.70 & 44$'$ 54$''.1$ & 16.45 & 17.24 & 17.00 & 16.79 & 16.43\\
4  & 36$^s$.62 & 45$'$ 09$''.2$ & 16.65 & 16.91 & 16.47 & 16.03 & 15.50\\
5  & 36$^s$.45 & 45$'$ 05$''.0$ & 16.64 & 16.92 & 16.47 & 16.02 & 15.48\\
6  & 37$^s$.46 & 45$'$ 11$''.7$ & 16.89 & 17.31 & 17.10 & 16.79 & 16.31\\
7  & 34$^s$.66 & 45$'$ 16$''.3$ & 16.98 & 17.27 & 16.84 & 16.42 & 15.82\\
8  & 33$^s$.96 & 45$'$ 10$''.2$ & 16.73 & 17.05 & 16.62 & 16.20 & 15.62\\
9  & 36$^s$.62 & 45$'$ 18$''.6$ & 17.09 & 17.24 & 16.79 & 16.36 & 15.84\\
10 & 36$^s$.44 & 45$'$ 15$''.3$ & 17.16 & 17.24 & 16.72 & 16.26 & 15.67\\
11 & 35$^s$.68 & 45$'$ 22$''.2$ & 17.58 & 17.71 & 17.24 & 16.77 & 16.18\\
12 & 36$^s$.10 & 45$'$ 28$''.4$ & 17.42 & 17.63 & 17.22 & 16.83 & 16.27\\
13 & 35$^s$.73 & 45$'$ 28$''.0$ & 17.56 & 17.72 & 17.31 & 16.91 & 16.36\\
14 & 33$^s$.40 & 45$'$ 20$''.8$ & 17.51 & 17.83 & 17.48 & 17.08 & 16.51\\
15 & 32$^s$.84 & 45$'$ 13$''.2$ & 17.82 & 17.93 & 17.49 & 17.06 & 16.43\\
16 & 32$^s$.29 & 45$'$ 05$''.2$ & 17.75 & 18.02 & 17.68 & 17.30 & 16.71\\
17 & 35$^s$.67 & 44$'$ 39$''.9$ & 17.22 & 17.49 & 17.09 & 16.69 & 16.14\\
18 & 37$^s$.91 & 45$'$ 03$''.3$ & 18.58 & 19.09 & 18.70 & 18.29 & 17.64\\
19 & 37$^s$.47 & 44$'$ 29$''.8$ & 18.05 & 18.53 & 18.33 & 17.93 & 17.48\\
20 & 36$^s$.47 & 44$'$ 26$''.0$ & 17.52 & 17.90 & 17.49 & 17.08 & 16.58\\
21 & 36$^s$.02 & 44$'$ 27$''.1$ & 17.76 & 18.02 & 17.57 & 17.15 & 16.60\\
22 & 36$^s$.03 & 44$'$ 24$''.3$ & 17.32 & 17.83 & 17.47 & 17.07 & 16.63\\
23 & 35$^s$.20 & 44$'$ 23$''.6$ & 17.42 & 17.66 & 17.32 & 16.95 & 16.43\\
24 & 35$^s$.30 & 44$'$ 23$''.7$ & 17.38 & 17.62 & 17.30 & 16.93 & 16.42\\
25 & 34$^s$.17 & 44$'$ 25$''.9$ & 17.57 & 17.84 & 17.43 & 17.06 & 16.50\\
26 & 33$^s$.71 & 44$'$ 26$''.6$ & 17.66 & 17.94 & 17.59 & 17.22 & 16.68\\
27 & 33$^s$.42 & 44$'$ 38$''.6$ & 17.35 & 17.62 & 17.25 & 16.87 & 16.28\\
28 & 33$^s$.40 & 44$'$ 44$''.5$ & 17.54 & 17.67 & 17.24 & 16.81 & 16.21\\
29 & 33$^s$.31 & 44$'~48''.2$   & 17.45 & 17.59 & 17.15 & 16.75 & 16.12\\
30 & 32$^s.84$ & 44$'~45''.7$   & 17.53 & 17.88 & 17.50 & 17.15 & 16.54\\
31 & 31$^s.68$ & 45$'~07''.7$   & 17.78 & 18.26 & 17.99 & 17.66 & 17.10\\
32 & 32$^s.06$ & 45$'~20''.3$   & 17.45 & 17.89 & 17.58 & 17.27 & 16.79\\
33 & 32$^s$.16 & 45$'~22''.4$   & 17.43 & 17.87 & 17.58 & 17.27 & 16.80\\
34 & 31$^s$.93 & 45$'~24''.3$   & 17.47 & 17.91 & 17.68 & 17.36 & 16.88\\
35 & 32$^s$.02 & 45$'~30''.7$   & 17.49 & 18.04 & 17.78 & 17.46 & 17.02\\
36 & 35$^s$.45 & 45$'~14''.6$   & 17.61 & 17.57 & 16.91 & 16.43 & 15.76\\
37\tablenotemark{3} & 35$^s$.68 & 45$'~06''.2$  & 17.61 & 17.25  & 16.40  & 15.84 & 15.20\\
38\tablenotemark{4} & 34$^s$.15 & 45$'~00''.5$  & 17.50 & 17.29  & 16.48  & 15.93 & 15.25\\
39 & 34$^s$.69 & 45$'~50''.0$   & 18.69 & 18.96 & 18.63 & 18.24 & 17.75\\
40 & 33$^s$.63 & 45$'~51''.4$   & 18.35 & 18.72 & 18.45 & 18.12 & 17.70\\
41 & 36$^s$.32 & 45$'~42''.3$   & 18.16 & 18.31 & 17.96 & 17.62 & 17.12\\
42\tablenotemark{5} & 36$^s$.74 & 45$'~32''.9$ & 18.14 & 18.29  & 17.85  & 17.42 & 16.91\\
43 & 37$^s$.03 & 45$'~28''.1$   & 17.88 & 18.16 & 17.72 & 17.28 & 16.77\\
44 & 37$^s$.39 & 45$'~35''.4$   & 18.14 & 18.32 & 17.99 & 17.62 & 17.09\\
45 & 34$^s$.42 & 45$'~31''.9$   & 17.73 & 17.83 & 17.45 & 17.07 & 16.52\\
46 & 33$^s$.23 & 45$'~26''.9$   & 17.64 & 17.82 & 17.41 & 17.02 & 16.48\\
47 & 35$^s$.01 & 44$'~47''.3$   & 17.67 & 17.62 & 17.05 & 16.59 & 15.94\\
48 & 37$^s$.02 & 44$'~59''.1$   & 17.58 & 17.86 & 17.44 & 17.00 & 16.46\\
49 & 36$^s$.37 & 44$'~48''.5$   & 17.82 & 17.88 & 17.37 & 16.90 & 16.31\\
50 & 35$^s$.62 & 44$'~49''.1$   & 17.65 & 17.75 & 17.20 & 16.72 & 16.09\\
51 & 35$^s$.82 & 44$'~54''.9$   & 17.38 & 17.49 & 16.97 & 16.52 & 15.89\\
52 & 31$^s$.03 & 44$'~49''.6$   & 19.42 & 19.60 & 19.01 & 18.58 & 17.79\\
53 & 32$^s$.06 & 44$'~31''.5$   & 19.21 & 19.67 & 19.05 & 18.61 & 17.81\\
54 & 38$^s$.19 & 44$'~38''.1$  & 19.04 & 19.34 & 19.01 & 18.61 & 17.92\\
55 & 38$^s$.32 & 44$'~44''.2$  & 19.32 & 19.56 & 19.12 & 18.66 & 18.02\\
56 & 36$^s$.95 & 44$'~37''.0$  & 18.51 & 18.63 & 18.18 & 17.76 & 17.13\\
57 & 32$^s.44$ & 44$'~47''.7$  & 17.96 & 18.23 & 17.83 & 17.42 & 16.84\\
58 & 36$^s.22$ & 44$'~09''.0$  & 19.70 & 19.93 & 19.48 & 19.07 & 18.30\\
59 & 36$^s.60$ & 44$'~10''.4$  & 19.42 & 19.81 & 19.43 & 19.00 & 18.29\\
60 & 37$^s.12$ & 45$'~44''.9$  & 18.90 & 19.02 & 18.71 & 18.28 & 17.77\\
61 & 37$^s.55$ & 45$'~44''.3$  & 18.62 & 18.84 & 18.63 & 18.23 & 17.86\\
62 & 35$^s.56$ & 46$'~04''.5$  & 19.68 & 18.84 & 18.63 & 18.23 & 17.86\\
63 & 34$^s.39$ & 46$'~04''.5$  & 19.61 & 19.80 & 19.63 & 19.27 & 18.96\\
64 & 33$^s.87$ & 46$'~06''.0$  & 19.81 & 19.90 & 19.85 & 19.36 & 19.12\\
65 & 32$^s.70$ & 45$'~53''.1$  & 19.50 & 19.69 & 19.46 & 19.18 & 18.70\\
66 & 39$^s.05$ & 44$'~52''.2$  & 19.57 & 19.87 & 19.32 & 18.96 & 18.21\\
67 & 39$^s.39$ & 45$'~04''.5$  & 20.07 & 20.12 & 19.63 & 19.08 & 18.49\\
68 & 39$^s.47$ & 45$'~08''.3$  & 20.10 & 20.23 & 19.86 & 19.29 & 18.60\\
69 & 38$^s.72$ & 45$'~01''.7$  & 19.96 & 20.07 & 19.56 & 19.16 & 18.18\\
70 & 39$^s.10$ & 44$'~53''.7$  & 19.62 & 19.85 & 19.35 & 18.97 & 18.23\\
71 & 31$^s.10$ & 44$'~18''.4$  & 20.42 & 20.70 & 20.31 & 19.89 & 18.76\\
72 & 30$^s.98$ & 44$'~13''.9$  & 20.51 & 20.92 & 20.48 & 20.17 & 18.88\\
73 & 31$^s.07$ & 44$'~07''.1$  & 20.14 & 21.24 & 20.66 & 20.33 & 19.16\\
74 & 30$^s.30$ & 45$'~19''.5$  & 20.61 & 21.01 & 21.34 & 20.58 & 19.43\\
75 & 43$^s.42$ & 46$'~10''.0$  & 20.90 & 21.30 & 20.79 & 20.02 & 20.73\\
76 & 36$^s.69$ & 45$'~08''.2$  & 16.73 & 16.99 & 16.55 & 16.12 & 15.62\\
77 & 35$^s.94$ & 45$'~10''.6$  & 17.46 & 17.31 & 16.59 & 16.09 & 15.44\\
78 & 33$^s.99$ & 44$'~46''.8$  & 18.01 & 18.02 & 17.36 & 16.84 & 16.15\\
79 & 36$^s.44$ & 45$'~38''.6$  & 18.25 & 18.28 & 17.93 & 17.56 & 17.04\\
80 & 39$^s.07$ & 45$'~05''.7$  & 20.31 & 20.21 & 19.53 & 19.13 & 18.27\\
81 & 35$^s.07$ & 45$'~21''.7$  & 18.27 & 18.20 & 17.60 & 17.12 & 16.41\\
\enddata
\tablenotetext{1}{Centermost region (nucleus + part of bulge) of the galaxy;} 
\tablenotetext{2}{Positions of peak of emission, in afilter where the intensity was strongest,
$\alpha$(J2000.0)=$10^h~46^m$; $\delta$(J2000.0)=$+13^{\circ}$;}
\tablenotetext{3}{northeast region of the stellar bar;}
\tablenotetext{4}{southwest region of the stellar bar;}
\tablenotetext{5}{position of peak which includes supernova 2003AA type Ia \citep{swi03,fil03}.}
\end{deluxetable}

\begin{deluxetable}{rrrrrrrr}
\tabletypesize{\tiny}
\tablecolumns{8} \tablewidth{0pc}
\tablecaption{$U B V R I$ Photometry of stars and a background galaxy in 
the field of NGC 3367, corrected for Galactic extinction, using IRAF task {\it phot} using an aperture with radius 8 pixels $\sim3''.41\sim715$ pc} 
\tablehead{ \colhead{ID} & \colhead{$\alpha$(J2000.0)} & \colhead{$\delta$(J2000.0)} &
\colhead{U} & \colhead{B} & \colhead{V} & \colhead{R} & \colhead{I}} 
\startdata
82\tablenotemark{1} & 33$^s.10$ & 45$'~11''.4$ & 17.91 & 17.94 & 17.37 & 16.86 & 16.21\\ 
83                  & 28$^s.56$ & 45$'~54''.7$ & 19.66 & 19.04 & 18.13 & 17.52 & 17.04\\
84                  & 42$^s.53$ & 44$'~33''.5$ & 18.20 & 17.94 & 17.19 & 16.75 & 16.36\\
85\tablenotemark{2} & 27$^s.56$ & 43$'~46''.0$ & 13.93 & 13.54 & 12.91 & 12.51 & 12.26\\
86 & 29$^s.34$ & 44$'~12''.5$ & 21.85 & 21.73 & 20.29 & 19.40 & 17.93\\
87 & 29$^s.78$ & 43$'~27''.9$ & 21.64 & 20.99 & 20.09 & 19.29 & 18.45\\
88 & 39$^s.28$ & 44$'~08''.0$ & 20.63 & 20.34 & 19.63 & 18.65 & 17.32\\
89\tablenotemark{3} & 28$^s.62$ & 42$'~41''.5$ & 20.63 & 20.34 & 19.13 & 18.43 & 17.56\\
90\tablenotemark{4} & 30$^s.60$ & 42$'~28''.6$ & 13.16 & 13.03 & 12.73 & 12.41 & 12.17\\
\enddata
\tablenotetext{1}{dim star within NGC 3367;} 
\tablenotetext{2}{bright star south west in the field of NGC 3367; this star was taken as a test star in order to estimate the uncertainities when obtaining its magnitudes with different aperture radii;} 
\tablenotetext{3}{small galaxy with weak surface brightness to the south south west 
in the field of NGC 3367;} 
\tablenotetext{4}{bright star south south west in the field of NGC 3367; this star was taken as a test star in order to estimate the uncertainities when obtaining its magnitudes with different aperture radii.}

\end{deluxetable}

\begin{deluxetable}{rrrrr}
\tablecolumns{5} \tablewidth{0pc}
\tabletypesize{\tiny}
\tablecaption{U-B, B-V, V-I Colors of Stellar Structures in NGC 3367.}
\tablehead{\colhead{ID} & \colhead{$U - B$} & \colhead{$B - V$} & \colhead{$V
- I$} & \colhead{SWB type\tablenotemark{1}}}
\startdata
1\tablenotemark{2}  & -0.55 & 0.47 & 1.00 & I\\
2  & -0.79 & 0.23 & 0.60 & I\\
3  & -0.79 & 0.24 & 0.57 & I\\
4  & -0.26 & 0.44 & 0.97 & I\\
5  & -0.28 & 0.45 & 0.99 & I\\
6  & -0.42 & 0.21 & 0.79 & I\\
7  & -0.29 & 0.43 & 1.02 & I\\
8  & -0.32 & 0.43 & 1.00 & I\\
9  & -0.15 & 0.45 & 0.95 & II\\
10 & -0.08 & 0.52 & 1.05 & III\\
11 & -0.13 & 0.47 & 1.06 & II\\
12 & -0.21 & 0.41 & 0.95 & II\\
13 & -0.16 & 0.41 & 0.95 & II\\
14 & -0.32 & 0.35 & 0.97 & I\\
15 & -0.11 & 0.44 & 1.06 & III\\
16 & -0.27 & 0.34 & 0.97 & I\\
17 & -0.27 & 0.40 & 0.95 & I\\
18 & -0.51 & 0.39 & 1.05 & I\\
19 & -0.48 & 0.20 & 0.84 & I\\
20 & -0.38 & 0.41 & 0.91 & I\\
21 & -0.26 & 0.51 & 0.97 & I\\
22 & -0.51 & 0.36 & 0.84 & I\\
23 & -0.24 & 0.34 & 0.89 & I\\
24 & -0.24 & 0.32 & 0.88 & I\\
25 & -0.27 & 0.41 & 0.93 & I\\
26 & -0.28 & 0.35 & 0.91 & I\\
27 & -0.27 & 0.37 & 0.97 & I\\
28 & -0.13 & 0.43 & 1.03 & II\\
29 & -0.14 & 0.44 & 1.03 & II\\
30 & -0.35 & 0.38 & 0.96 & I\\ 
31 & -0.48 & 0.27 & 0.89 & I\\
32 & -0.44 & 0.31 & 0.79 & I\\
33 & -0.44 & 0.29 & 0.78 & I\\
34 & -0.44 & 0.23 & 0.80 & I\\
35 & -0.55 & 0.26 & 0.76 & I\\
36 & +0.04 & 0.66 & 1.15 & VII\\
37\tablenotemark{3} & +0.36 & 0.85 & 1.20 & VI\\
38\tablenotemark{4} & +0.21 & 0.81 & 1.23 & VII\\
39 & -0.27 & 0.33 & 0.88 & I\\
40 & -0.37 & 0.27 & 0.75 & I\\
41 & -0.15 & 0.35 & 0.84 & II\\
42\tablenotemark{5} & -0.15 & 0.44 & 0.94 & II\\
43 & -0.28 & 0.44 & 0.95 & I\\
44 & -0.18 & 0.33 & 0.88 & II\\
45 & -0.10 & 0.38 & 0.93 & III\\
46 & -0.18 & 0.41 & 0.93 & II\\
47 & +0.05 & 0.57 & 1.11 & IVB\\
48 & -0.28 & 0.42 & 0.98 & I\\
49 & -0.06 & 0.51 & 1.06 & IVA\\
50 & -0.10 & 0.55 & 1.11 & III\\
51 & -0.11 & 0.52 & 1.08 & III\\
52 & -0.18 & 0.59 & 1.22 & I\\
53 & -0.44 & 0.60 & 1.24 & I\\
54 & -0.30 & 0.33 & 1.09 & I\\
55 & -0.24 & 0.44 & 1.10 & I\\
56 & -0.12 & 0.45 & 1.05 & III\\
57 & -0.27 & 0.40 & 0.99 & I\\
58 & -0.23 & 0.45 & 1.18 & I\\
59 & -0.39 & 0.38 & 1.14 & I\\
60 & -0.12 & 0.31 & 0.94 & III\\
61 & -0.22 & 0.21 & 0.77 & II\\
62 & -0.01 & 0.17 & 0.75 & III\\
63 & -0.19 & 0.17 & 0.67 & II\\
64 & -0.09 & 0.05 & 0.73 & III\\
65 & -0.19 & 0.23 & 0.76 & II\\
66 & -0.30 & 0.55 & 1.11 & I\\
67 & -0.05 & 0.49 & 1.14 & IVA\\
68 & -0.13 & 0.37 & 1.26 & III\\
69 & -0.11 & 0.51 & 1.38 & III\\
70 & -0.23 & 0.50 & 1.12 & I\\
71 & -0.28 & 0.38 & 1.55 & I\\
72 & -0.41 & 0.44 & 1.60 & I\\
73 & -1.10 & 0.58 & 1.50 & I\\
74 & -0.40 & -0.33 & 1.91 & II\\
75 & -0.40 & 0.51 & 0.36 & I\\
76 & -0.26 & 0.44 & 0.94 & I\\
77 & +0.15 & 0.72 & 1.15 & VII\\
78 & -0.01 & 0.66 & 1.21 & VII\\
79 & -0.03 & 0.35 & 0.89 & IVA\\
80 & +0.10 & 0.68 & 1.26 & VII\\
81 & +0.07 & 0.60 & 1.19 & IVB\\
\enddata
\tablenotetext{1}{SWB type \citep{sea80,bic96}. Each type suggest
an approximate interval of age to different SWB classes as
follows: SWB 0 have ages between 0 and $10^7$ yrs, SWB I have ages
between $10^7$ and $3\times10^7$ yrs, SWB II have ages between
$3\times10^7$ and $7\times10^7$ yrs, SWB III have ages between
$7\times10^7$ and $2\times10^8$ yrs, SWB IVA have ages between
$2\times10^8$ and $4\times10^8$ yrs, SWB IVB have ages between
$4\times10^8$ and $8\times10^8$ yrs, SWB V have ages between
$8\times10^8$ and $2\times10^9$ yrs, SWB VI have ages between
$2\times10^9$ and $5\times10^9$ yrs, and SWB VII have ages between
$5\times10^9$ and $1.6\times10^{10}$ yrs;}
\tablenotetext{2}{Centermost region (nucleus + part of bulge) of
the galaxy;} 
\tablenotetext{3}{northeast region of the stellar bar;} 
\tablenotetext{4}{southwest region of the stellar bar;}
\tablenotetext{5}{position of peak which includes supernova 2003AA type Ia \cite{swi03,fil03};}  
\end{deluxetable}


\begin{deluxetable}{rrrrr}
\tablecolumns{5} \tablewidth{0pc}
\tabletypesize{\tiny}
\tablecaption{UBVI Colors of foreground stars and a background galaxy in the field of NGC 3367.}
\tablehead{\colhead{ID} & \colhead{$U - B$} & \colhead{$B - V$} & \colhead{$V- I$} & 
\colhead{type\tablenotemark{1}}}
\startdata
82\tablenotemark{2} & -0.03 & +0.57 & 1.16 & F7V\\
83                  & +0.62 & +0.91 & 1.09 & K3V\\ 
84                  & +0.26 & +0.75 & 0.83 & G8V\\
85\tablenotemark{3} & +0.39 & +0.63 & 0.65 & G2III\\
86                  & -0.88 & +1.44 & 2.36 & K5III \\
87                  & +0.65 & +0.90 & 1.64 & G7III \\
88                  & +0.29 & +0.71 & 2.31 & K3III \\
89\tablenotemark{4} & +0.29 & +1.21 & 1.57 & spiral\\
90\tablenotemark{5} & +0.13 & +0.30 & 0.56 & F1III\\
\enddata
\tablenotetext{1} {approximate spectral type \cite{mih81};}
\tablenotetext{2} {dim star within NGC 3367;}\\
\tablenotetext{3}{bright star south west in the field of NGC 3367; this star was taken as a test star in order to estimate the uncertainities when obtaining its magnitudes with different aperture radii;} 
\tablenotetext{4}{galaxy with weak surface brightness south south west in the field of NGC 3367;}
\tablenotetext{5}{bright star south south west in the field of NGC 3367; this star was taken as a test star in order to estimate the uncertainities when obtaining its magnitudes with different aperture radii.}  
\end{deluxetable}

\clearpage
\newpage

\end{document}